%% file: main.tex
\documentclass[a4paper,11pt]{article}
\pdfoutput=1 
\usepackage{jinstpub} 
\usepackage{slantsc}
\usepackage{subcaption}
\usepackage{textalpha}
\usepackage[table]{xcolor}
\usepackage{multirow}
\newcommand{\AmLi}[0]{AmLi \hspace{0.5pt}}
\newcommand{\AmBe}[0]{AmBe \hspace{0.5pt}}

\newcommand{\Am}[0]{\textsuperscript{241}Am \hspace{0.5pt}}
\newcommand{\Li}[0]{\textsuperscript{7}Li \hspace{0.5pt}}

\title{Design and characterization of AmLi neutron sources for the LZ experiment}
\author[]{A.B.M.R.~Sazzad,}
\author[]{J.~Busenitz,}
\author[]{ A.~Piepke,}
\author[1]{ S.~Poudel,\note{Now at ASML.}}
\author[]{ H.~Trewin,}
\author[2]{and A.~LeViness\note{Now at Princeton Plasma Physics Laboratory, 100 Stellarator Road, Princeton, NJ 08540.} }

\affiliation[]{Department of Physics and Astronomy, University of Alabama, Tuscaloosa, USA}
\emailAdd{asazzad@crimson.ua.edu}

\date{\today}
\abstract{In this paper we describe the development, testing, and characterization of three low-emission rate \AmLi neutron sources. The sources are used to calibrate the nuclear recoil response of the LUX-ZEPLIN (LZ) dark matter experiment. The sources' neutron emission rate was measured using \textsuperscript{3}He proportional tubes. The sources' gamma emissions were characterized using a high-purity germanium (HPGe) detector. Source-validated \textsc{Geant4} Monte Carlo simulations allowed to calibrate the Ge and neutron detectors' responses.}

\begin{document}
\maketitle
\flushbottom

\section{Introduction\label{sec:intro}}
\input{introduction}

\section{Design and manufacture\label{sec:design}}
\input{design_and_manufacture}

\section{Leak testing of the AmLi sources\label{sec:leak}}
\input{leak_testing}

\section{Source activity \label{source_activity}}
\input{source_activity}

\section{Source \texorpdfstring{$\gamma$} --- ray emission rate and spectrum \label{sec:gammaBkg}}

\input{g_fluence}

\section{Neutron emission rate measurement\label{sec:n_fluence}}
\input{n_fluence_calibration}

\section{Conclusion}
\input{Conclusion}

\acknowledgments
This project was conceived as a contribution to the LZ experiment. We would like to thank our LZ colleagues for their support and valuable insights. We acknowledge Jaclyn Schillinger's contribution in the early stages of this project.  This research was supported in part by the U.S. Department of Energy under DOE Grant DE-SC0012447 and by LZ project subcontract 7270503 from Lawrence Berkeley National Lab.

\input{bib4jinst}
\end{document}

%% file: introduction.tex
Neutron sources are widely used in various medical, industrial, and basic research applications. AmBe and $^{252}$Cf neutron sources are probably the most common. These two source types are  based on either ($\alpha$, n) nuclear reactions or spontaneous fission. The former type is a binary system, containing an $\alpha$-emitter as the driver and some low Z-material (Z denotes the atomic number) as the target for the intended reaction that results in the emission of neutrons after the formation of an excited compound nucleus. This paper describes the design, fabrication, and characterization of such a nuclear reaction based compound source using a mixture of americium and lithium. These AmLi neutron sources were developed for calibration of the LUX-ZEPLIN (LZ)~\cite{AKERIB2020163047} detector, searching for dark matter. 

The LZ experiment uses 10 tonnes of xenon and a dual phase time projection chamber (TPC) to search for elastic scattering of Weakly Interacting Massive Particles (WIMPs) off xenon nuclei.   The experimental signature of a WIMP interaction is the creation of scintillation light and ionization electrons in a particular proportion due to a recoiling xenon nucleus. The purpose of neutron sources is to calibrate the LZ detector response to nuclear recoils, in order to suppress the dominant background, which is due to electron recoils. Such background events yield different amounts of scintillation light and ionization electrons. Neutron source data helps reconstruct the recoil energy which, in turn, is related to the WIMP mass.  More information on the LZ experiment can be found in~\cite{AKERIB2020163047}. 

Low energy neutrons, with kinetic energies below 1.5 MeV, are well suited to calibrate the nuclear recoil (NR) energy response of xenon detectors. Events induced by low energy neutrons closely resemble the NR response to WIMP particles in the relevant energy range~\cite{LZTDR}. Such neutron source should emit as little $\gamma$-radiation as possible to limit unwanted background events in the calibration data. The development of the sources described in this paper paid particular attention to the suppression and quantification of source-related $\gamma$-radiation.

Based on these considerations, the choice fell on AmLi compound sources. This system emits neutrons with a maximum energy of 1.5 MeV; in particular, the NR endpoint energy of about 40 keV corresponds to the 1.5 MeV neutron energy endpoint provides an NR calibration point near the top of the LZ WIMP energy search window.  Moreover, unlike $^{252}$Cf and AmBe sources, neutrons emitted by the \AmLi system are not accompanied by reaction-correlated high-energy gammas. This absence of reaction-correlated gammas results in a sample of nuclear recoil events with relatively little electron recoil background. However, the reaction neutrons are superimposed to the $\gamma$-radiation emitted in the $^{241}$Am $\alpha$-decay. This source of background was carefully studied.

To the best of our knowledge, there are no current manufacturers of AmLi sources, and extant sources we have access to have neutron emission rates at the level of tens of kHz, far too high to be feasible for use in the LZ experiment. Such low-yield sources had to be developed and manufactured in our lab.  In this paper, we describe the design, fabrication, testing, and characterization of the three \AmLi sources to be used for LZ calibration purposes.

%% file: design_and_manufacture.tex
The design of the \AmLi sources was carried out with four goals in mind.  First, fabrication and use of the sources would be safe to personnel following reasonable procedures for radiation safety.  Second, in order to not overload the LZ data acquisition system (DAQ) system, sources with a combined neutron emission rate of order 100 n/s, when deployed simultaneously, were targeted, so that large calibration data sets could be acquired on the order of a day.   
Third, the \AmLi sources need to be clean in that the neutron emission spectra would not contain any contributions from ($\alpha$, n) interactions on beryllium, carbon, and oxygen, that might be present as contamination, as sometimes observed in commercial neutron sources~\cite{MOZHAYEV2021109472}.  Fourth, the copious gamma emission following $^{241}$Am $\alpha$-decay, especially at 59 keV, would be sufficiently attenuated to avoid pile--up effects and unwanted source-related $\gamma$-events.

 \AmLi is an ($\alpha$, n)-type source. Alphas from \Am decay bombard the \Li nuclei. In a few cases, this leads to a nuclear reaction of type \textsuperscript{7}Li$+\alpha \rightarrow$\textsuperscript{11}B\textsuperscript{*}. The excited $^{11}$B nucleus then decays to $^{10}$B via neutron emission.
 Two thresholds have been observed for the ($\alpha$, n) reaction in \AmLi : (i) at 4.379 MeV where the interaction leaves the \textsuperscript{10}B nucleus in its ground state and  (ii) at 5.51 MeV where the interaction leaves the \textsuperscript{10}B nucleus at an excited state of 720 keV~\cite{PhysRev.108.1025,VANDERZWAN1972615}. However, the production of neutrons through the latter is negligible because only $0.6\%$ of the \Am alphas have an energy of more than 5.51 MeV, even if they manage to reach a \Li nucleus without losing any energy in the source matrix material. The main contribution is from the former production channel as it results from the most abundant \Am alphas with 5.486 MeV with a branching ratio of 85\%. These alphas need to reach the \Li target without losing more than 1 MeV of kinetic energy. Due to these kinematics, the neutron production varies considerably among different \AmLi sources because of the dependence of the average $\alpha$-energy loss on the source geometry, encapsulation and shielding, degree of proximity of $^{241}$Am atoms to $^7$Li atoms achieved in mixing the americium and lithium materials, and presence of the impurities~\cite{MOORE2021164987}.  

Prior to embarking on the fabrication of AmLi sources, two prototypes were fabricated to gain experience in building such sources and to provide test beds for evaluating the efficacy of further shielding to absorb the copious \Am gammas.  The prototypes were sealed in stainless steel capsules with 1-mm wall thickness. The \Am activities and neutron emission rates were measured for these prototypes resulting in an average $n/\alpha$ yield of $(1.0\pm 0.1)\cdot 10^{-6}$.

\begin{figure}[ht!]
     \centering
     \begin{subfigure}[t]{0.32\textwidth}
         \centering
         \includegraphics[width=\textwidth, height= 5cm]{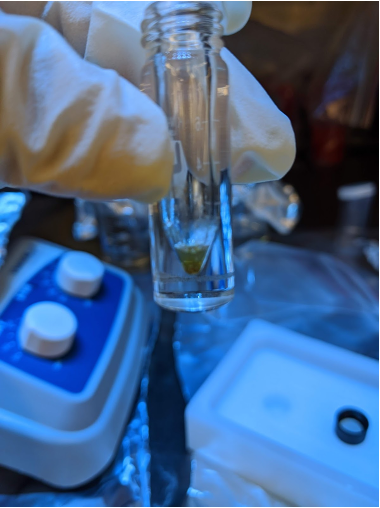}
         \label{fig:AmVial}
     \end{subfigure}
     \hfill
     \begin{subfigure}[t]{0.32\textwidth}
         \centering
         \includegraphics[width=\textwidth,height= 5cm]{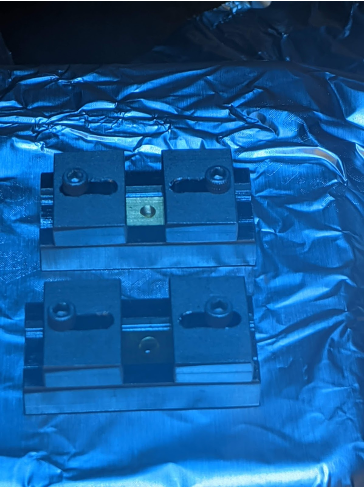}
         \label{fig:AmLiDep}
     \end{subfigure}
     \hfill
     \begin{subfigure}[t]{0.32\textwidth}
         \centering
         \includegraphics[width=\textwidth,height= 5cm]{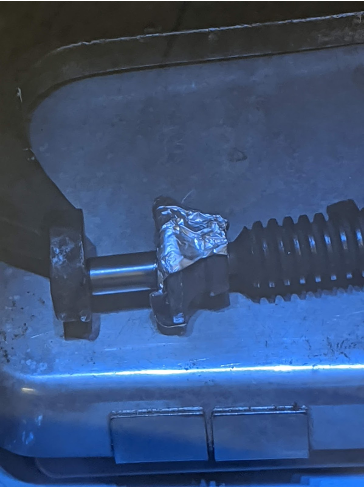}
         \label{fig:innerCapSeal}
     \end{subfigure}
        \caption{Left: \Am contained in a vendor-supplied vial in form of $\rm{Am(NO_3)_3}$. Each vial contains 1 mL of americium nitrate solution, which amounts to 37 MBq of \Am activity. Middle: \Am deposition in a gold foil. Right: Post-cure treatment of the inner capsule after its press-fit lid was sealed with epoxy.}
        \label{fig:sourceMatrix}
\end{figure}

\paragraph{\Am Deposition} 
Commercially available \AmLi sources typically contain rather high $^{241}$Am activities. 
The LZ DAQ~\cite{https://doi.org/10.48550/arxiv.1511.08385}, which is designed with the goal of saving maximal information for each triggered event,  cannot handle high event rates exceeding approximately 1000 events/s. Furthermore, since the commercial vendors often use the same facility to build different ($\alpha$, n) sources, impurities might be included~\cite{MOZHAYEV2021109472}. Hence, three sources were built at the University of Alabama (UA) as per LZ requirements. The \textsuperscript{241}Am was obtained from Eckert and Ziegler in the form of $\rm{Am(NO_3)_3}$ in 1 molar (M) nitric acid solution, shown in figure \ref{fig:sourceMatrix}.  
After slowly evaporating 1 mL of the manufacturer’s solution to near dryness and then reconstituting with 50 $\mu$L of 1 M nitric acid, the solution was deposited centrally on a gold foil of  0.1 mm thickness using a micro-liter syringe. The certificate of analysis for the gold foil specifies the purity of the gold foil to be better than 99.999\%. The only detected impurity is Ag, which is reported with a concentration of 1.8 ppm. $10\;\mu$L drops were deposited at a time on the gold foil until all the solution had been used up. The gold foil holder was placed on a hotplate inside a fume-hood and heated to evaporate the nitric acid solvent. The amount of $^{241}$Am deposited on the gold foil and not lost with the vapor during concentration and deposition, was estimated by counting the radiation emitted in the form of 59 keV gammas from \Am decay. More than 90\% of the \textsuperscript{241}Am was transferred to the gold foil in this process. 

In the next step, the \Am deposited gold foil was wrapped in a lithium foil. The task was carried out inside a glove box under an argon atmosphere to prevent the Li foil from oxidizing. The Li foil has a thickness of 0.75 mm. According to the certificate of analysis of the Li foil, it contains 99.9\% Li with some alkaline metals and nitrogen impurities.  The gold foil was laid on top of the lithium foil with the deposition side down, and then the combination was folded up ensuring good contact between them.
This Au-Am-Li sandwich constitutes the matrix of the AmLi source. 

\paragraph{Encapsulation of the source}
The encapsulation of the sources serves two purposes.  The first and more important is to ensure the source is safe to transport and handle following prescribed and accepted procedures. The encapsulation provides the barrier against the release of $^{241}$Am radioactivity into the environment. The second purpose is to absorb as much gamma radiation as feasible  from the source while at the same time not significantly affecting the neutron emission rate and energy spectrum. The outside dimensions of the sources were developed based on the geometrical and mass requirements defined by the LZ source deployment system.

The encapsulation consists of three nested metal capsules. The inner two are made of tungsten alloy (Midwest Tungsten Service MT-185), and the outer one is made of series 304 stainless steel (SS).  Each capsule is independently plugged and sealed. The inner capsule is closed by a press-fit plug and sealed with an acid-resistant epoxy (EP21ARHT from Masterbond), the middle capsule is closed by a threaded plug and sealed with the acid-resistant epoxy, and the outer capsule is closed by a press-fit plug and sealed by welding.  The dimensions of each capsule were chosen so that nesting was as tight as possible, allowing for machining tolerances and avoiding undue thermal stresses between capsules over the temperature range from -50 \textdegree C to 204 \textdegree C, which is the epoxy seal working range for the inner two capsules. 

To ensure that the internal temperature of the sources wouldn't exceed the working range of the epoxy during welding, careful testing was performed ahead of conducting the work.
To achieve low internal temperatures, TIG (tungsten inert gas) welding was
performed around the plug-body interface, a fraction of the arc at a time, followed by a cooling period.
Furthermore, the source bottom was placed into a heat sink during welding.
The heat sink was coupled to a water bath.
The suitability of this technique was tested before attempting to weld the high-activity sources using 8 outer source capsules, containing different temperature-sensing crayons (Tempilstik from Markal) instead of the tungsten bottles. The correctness of the melting temperature, stated by the manufacturer, was verified by means of putting crayon samples on a heating plate, equipped with a temperature probe.
The melting points of the crayons used during our welding tests ranged from  163 \textdegree C to  260 \textdegree C. 
The mock capsules were welded, following the procedure described above, and then cut open. None of the crayons was found to have melted, demonstrating that the internal temperature of the SS capsules was below  163 \textdegree C. 
This value is within the operating range of the epoxy used.  After completion of the welding, the weldments were required to pass a dye penetration test.

Furthermore, to test the integrity of the design, two prototypes containing approximately 5~$\mu$Ci (185 kBq) of \Am were subjected to thermal, vibration, pressure, impact, and puncture tests following guidance for neutron calibration sources from the International Organization for Standardization (ISO)-2919 standard~\cite{ISO2919} and then leak tested using the hot-acid method prescribed in the ISO-9978 standard~\cite{ISO9978}.  The activity found in the soaking liquid was consistent with the background.

%% file: leak_testing.tex
\begin{figure}[ht!]
     \centering
     \begin{subfigure}[t]{0.32\textwidth}
         \centering
         \includegraphics[width=\textwidth, height= 5cm]{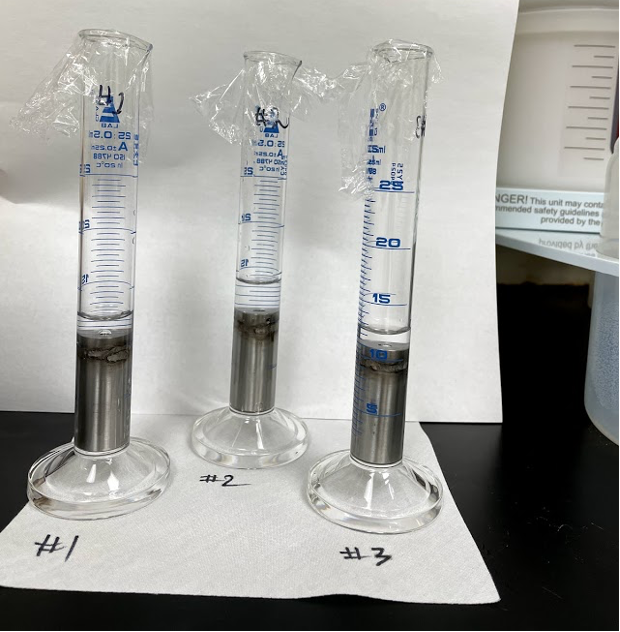}
         \label{fig:CapsuleSoak}
     \end{subfigure}
     \hfill
     \begin{subfigure}[t]{0.32\textwidth}
         \centering
         \includegraphics[width=\textwidth,height= 5cm]{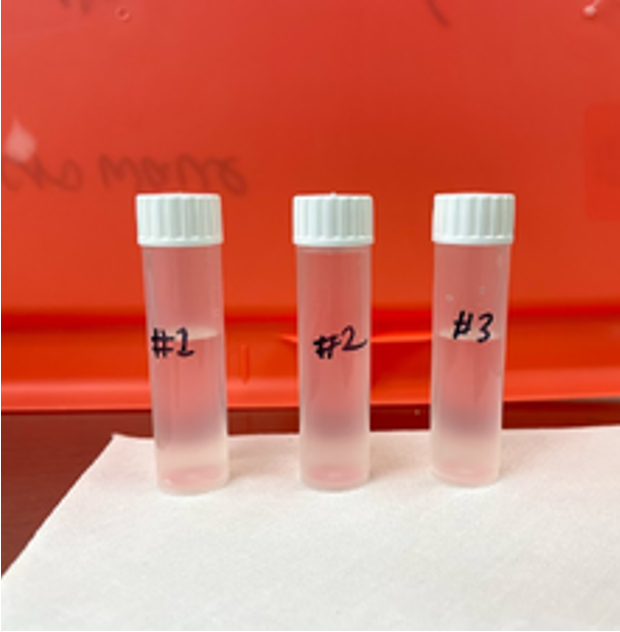}
         \label{fig:acidPEvials}
     \end{subfigure}
     \hfill
     \begin{subfigure}[t]{0.32\textwidth}
         \centering
         \includegraphics[width=\textwidth,height= 5cm]{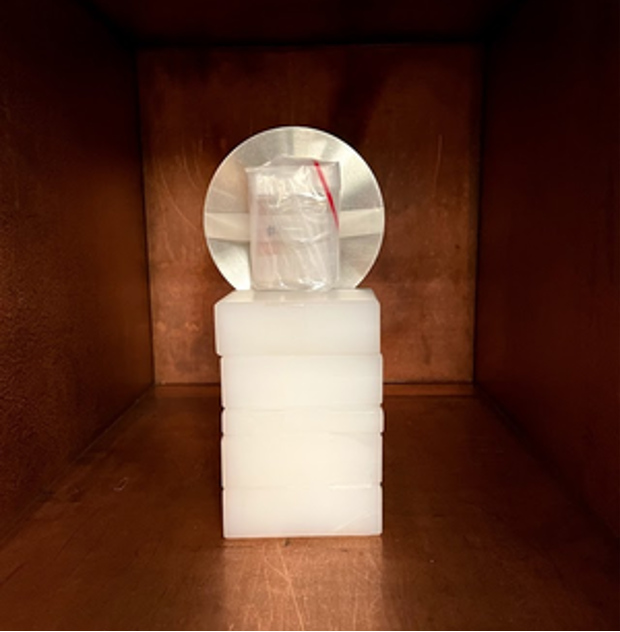}
         \label{fig:vialsGeDet}
     \end{subfigure}
        \caption{Left: Leak Testing of the sources. AmLi capsules submerged in 1 M nitric acid. Middle: Soak acid transferred to PE counting vials. Right: PE vials being counted in the GeIII setup.}
        \label{fig:LeakTest}
\end{figure}

Source integrity is of paramount importance during the calibration of ultra-low background experiments. Hence, the completed \AmLi sources underwent acid immersion and water pressurisation leakage testing as per the ISO-9978 standard. 
In order to perform the leak test, the sources were individually submerged in 7.5 ml of 1 M nitric acid for 24 hours. The soak acid was collected in small polyethylene (PE) vials, as shown in figure \ref{fig:LeakTest}, and counted using the GeIII, a low background radiation detection setup operated by our group. $73.2\%$ of the entire soak acid was transferred to the PE counting vials.

\begin{figure}[ht!]
     \centering
     \begin{subfigure}[t]{0.7\textwidth}
         \centering
         \includegraphics[width=\textwidth]{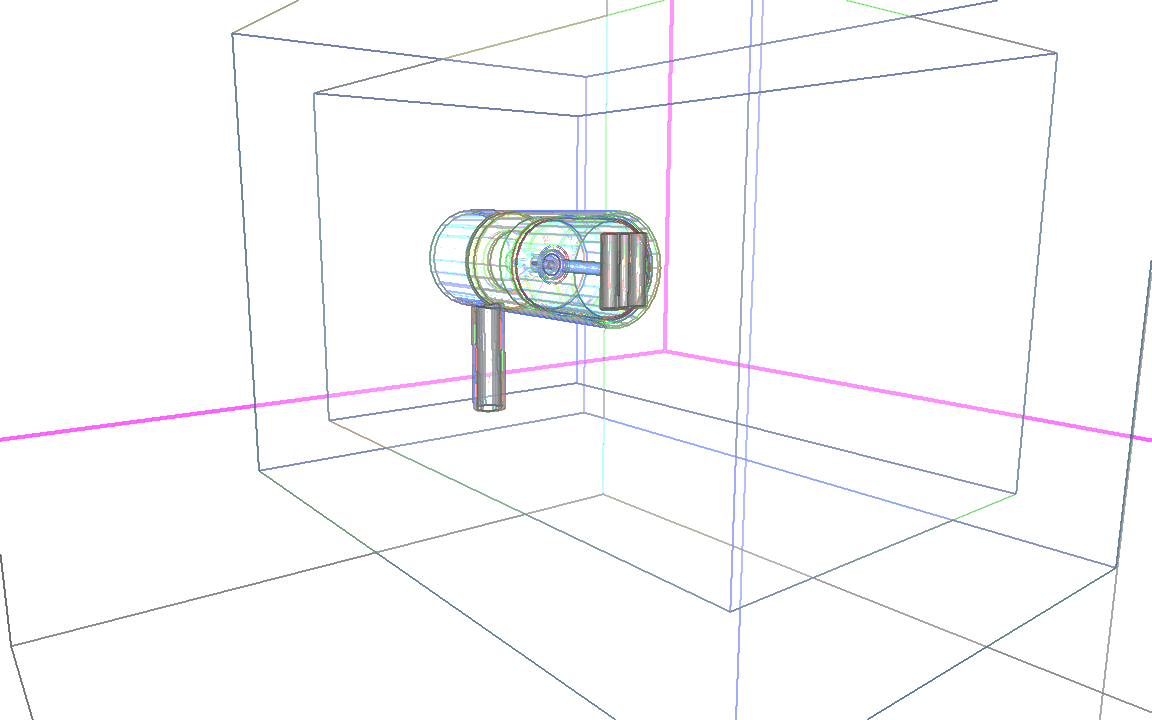}
         \label{fig:LeakTestMCside}
     \end{subfigure}
     \\
     \centering
     \begin{subfigure}[t]{0.8\textwidth}
         \centering
         \includegraphics[width=\textwidth]{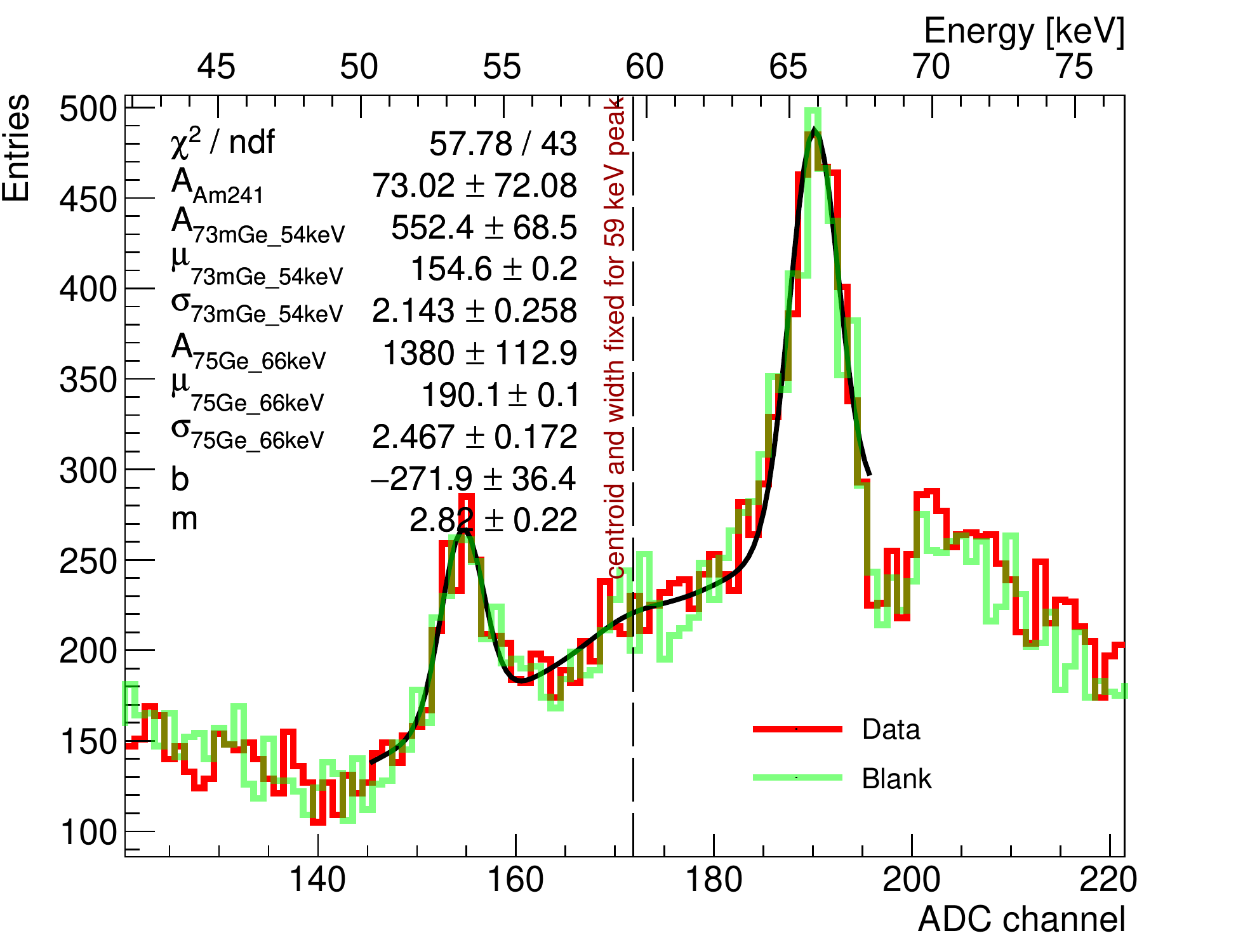}
         \label{fig:SoakDataAnalysis}
     \end{subfigure}
        \caption{Top: Leak test data analysis: view of the counting geometry as rendered by \textsc{Geant4}. Bottom: Soak data and blank data superimposed, the fitting of the 59 keV peak in the soak data shows no excess of events.}
        \label{fig:LeakTestDataAnalysis}
\end{figure}

 Any leak of \Am out of the capsule would be dissolved in the nitric acid (the original source solvent) and then be detected via the 59 keV full absorption peak in the GeIII $\gamma$-ray spectrum. No discernible 59 keV peak was observed. In order to perform a one-dimensional peak search in the resulting energy spectrum, the centroid and width of the hypothetical 59 keV peak were fixed to values derived from a direct calibration. To pinpoint the 59 keV peak parameters in terms of the ADC channel, an AmBe source was counted with GeIII. This AmBe source contains no gamma shielding, so the 59 keV peak could be observed with high statistics. Fit results from calibration were used to fix the peak nuisance parameters in the leak test data. This approach was used to put a limit on the \Am activity, as shown at the bottom of figure~\ref{fig:LeakTestDataAnalysis}. No statistically significant excess of peak events over a counting blank, consisting of acid-filled vials not used in the soak, was observed.

The detection efficiency of the used Ge detector was estimated by means of a \textsc{Geant4} Monte Carlo simulation, as described in~\cite{TSANG201975} (shown in the top of figure \ref{fig:LeakTestDataAnalysis}).  59 keV gammas were simulated originating from the liquid occupying the volume of the three vials. 
The simulated GeIII detector response was fitted with a Gaussian response, superimposed to a linear background. From this, the detection efficiency was estimated to be 0.0681 $\pm$ 0.0003\textsuperscript{stat} $\pm$ 0.0050\textsuperscript{syst} per primary gamma emitted. The systematic error estimation, done with 46 keV $\gamma$-radiation from \textsuperscript{210}Pb,  was taken from \cite{TSANG201975}. 
Furthermore, a blank sample was made from soaking two mock capsules which had not been exposed to \Am but were handled in the lab the same way the source capsules had. A third vial was filled with acid, handled the same way as the others but not containing a metal capsule and added to the suite of counting vials. The counting result from the soak data and the blank sample is shown at the bottom of Figure \ref{fig:LeakTestDataAnalysis}. The sample and blank were both counted for 12.6 days. The agreement between the spectra indicates no \Am is present in the soak acid. Subtracting the blank rate from the soak data count rate, dividing by the detection efficiency, and dividing by the appropriate branching ratio, a limit was placed on the \Am activity in the acid. Accounting for the fact that only $73.2\%$ of the soak acid was collected, the final activity limit set on the soak acid from all three sources was $< 5.0$  mBq at $90\%$ confidence level using the Feldman Cousins method \cite{FeldmanCousins}. When taken relative to the \Am activity contained in all three sources (discussed in the next section) this limit corresponds to a leakage fraction of $<5.3\cdot 10^{-11}$ of the contained activity; a strong constraint indeed. 

For the water pressurisation leakage test, all three sources were accurately weighed after cleaning their surfaces with methanol and acetone. They were submerged in de-ionized (DI) water in open beakers. These beakers were then placed into a pressure vessel and subjected to an air pressure of 3.9 atm (57.3 psi gauge pressure). The vessel was sealed and left pressurized for about 18 hours- at the end of that period, the pressure was dropped to 2.8 atm. Next, the sources were recovered, wiped, and rinsed with methanol and acetone to remove any remaining water. The sources were weighed again and were found $0.1 \sim 0.2$ mg lighter than their initial mass. The pressure test indicates that there were no detectable water leaks into the capsules. We interpreted this observation to mean that there are no penetrations through the welds.

%% file: source_activity.tex
\begin{figure}[ht!]
     \centering
     \begin{subfigure}[t]{0.35\textwidth}
         \centering
         \includegraphics[width=\textwidth]{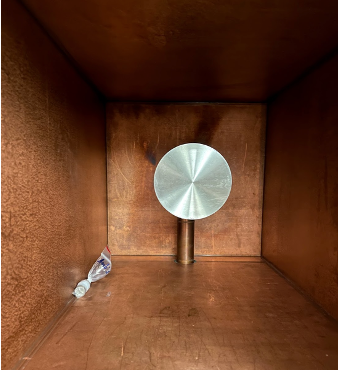}
     \end{subfigure}
     \hfill
     \begin{subfigure}[t]{0.55\textwidth}
         \centering
         \includegraphics[width=\textwidth]{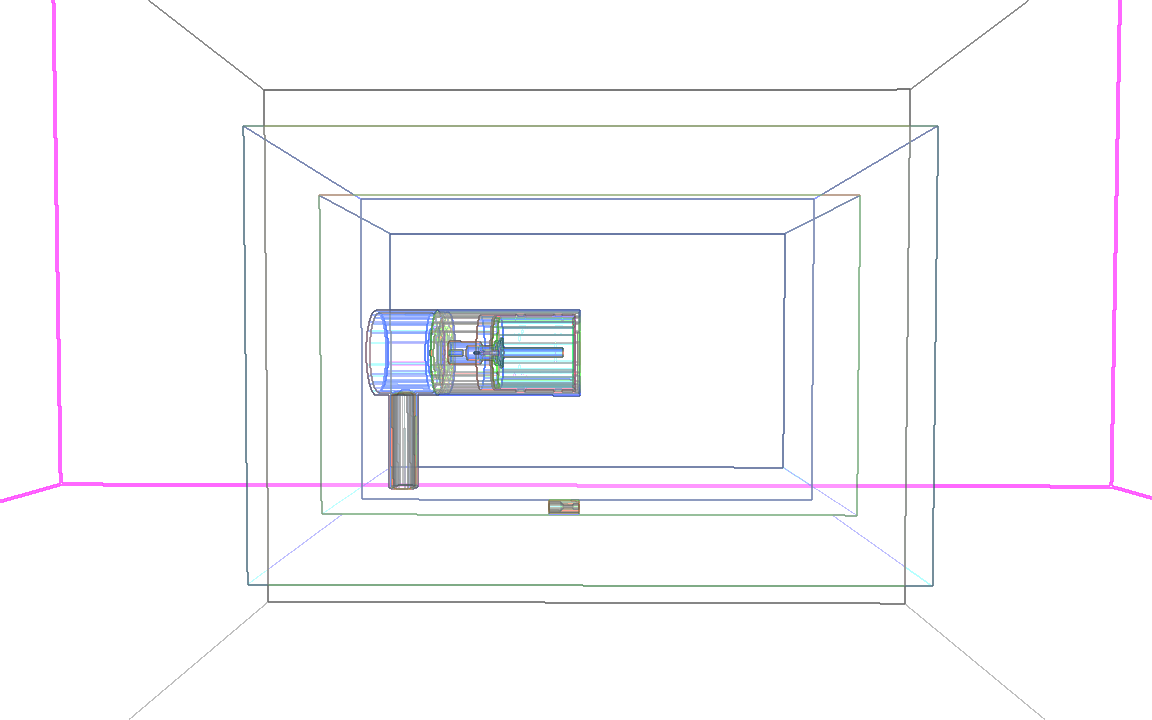}
     \end{subfigure}
             \caption{Left: Measurement of the \Am 
             activity contained in each source. The AmLi source was placed at the ``remote location'' of the detector to reduce solid angle uncertainty. Right: Side-view of the setup as rendered by the \textsc{Geant4} model.}
        \label{fig:MCforAmLiGammaMeas}
\end{figure}
\begin{figure}[ht!]
         \centering
         \includegraphics[width=0.7\textwidth]{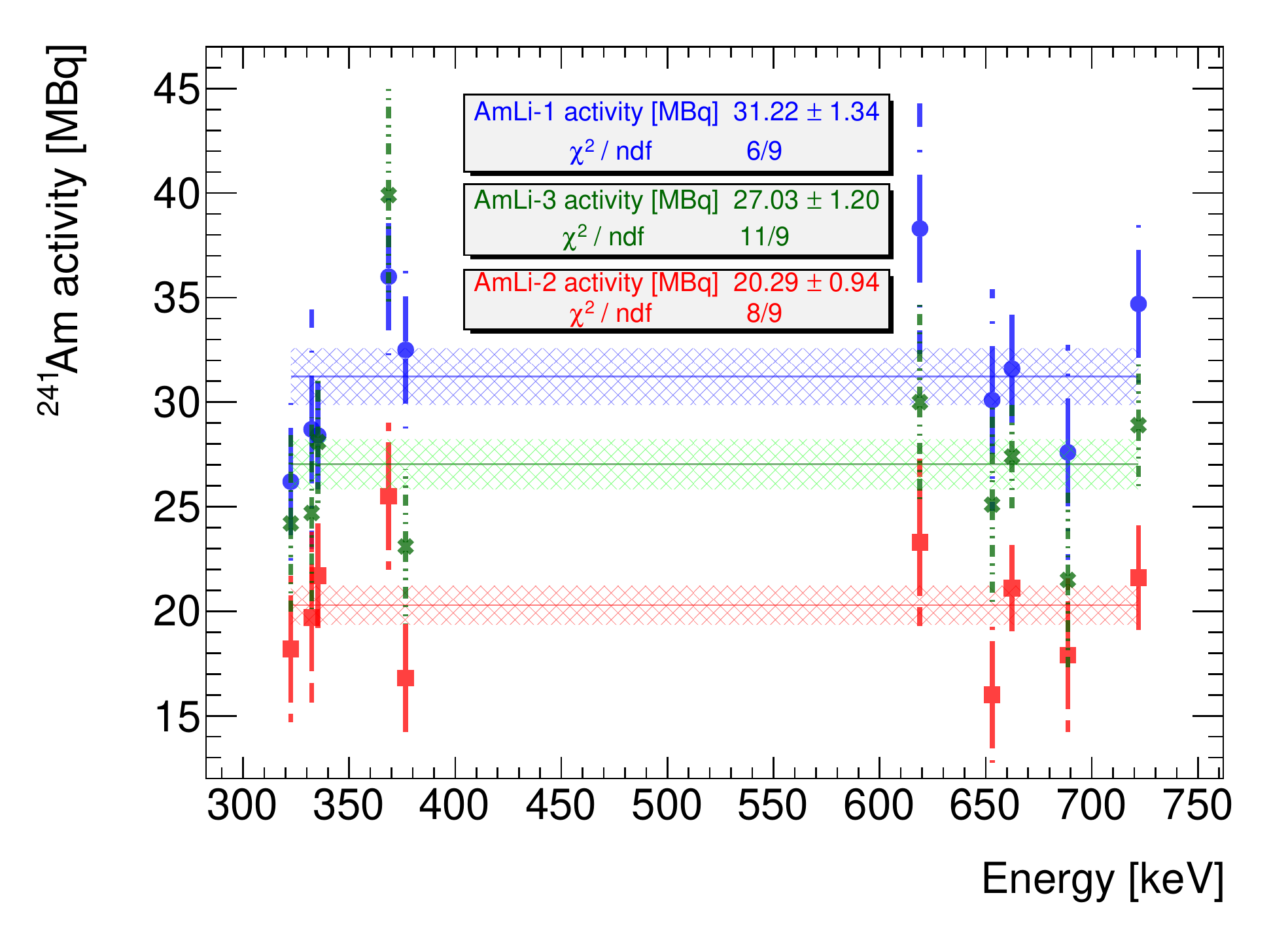}
            \caption{Best fit results for the \Am activity of all three sources. Blue circles, red squares, and green crosses denote the data sets for AmLi-1, AmLi-2, and AmLi-3 respectively. }
        \label{fig:Am241ContentEstimate}
\end{figure}

The \Am activity contained in each of the three sources was determined to quantify the $\gamma$-ray induced background as well as their $\gamma$-ray to neutron emissivity ratio. To obtain the \Am quantity contained in the sources, ten high energy $\gamma$-quanta co-emitted in the $^{241}$Am $\alpha$-decay  were measured with the GeIII setup. Unlike the low energy 59 keV $\gamma$-radiation which is significantly suppressed by the tungsten encapsulations of the source, these penetrating high energy gammas have enough statistics to be detected with our counting setup. The amount of \Am in the source matrix is then estimated from the peak integral returned from the fit result divided by the corresponding branching ratios and detection efficiencies derived from Monte Carlo simulation in \textsc{Geant4} as shown in figure~\ref{fig:MCforAmLiGammaMeas} (right). For each source, the ten peak-wise activities were then fitted with a constant to get the source activity as shown in figure~\ref{fig:Am241ContentEstimate}.  A systematic uncertainty of 6\% was added quadratically to the statistical uncertainty of each peak to get an acceptable chi-square value. We interpret this as an independent evaluation of the systematic uncertainty of the $\gamma$-ray detection efficiency correction. The estimated \Am activities of the three sources, along with their $\gamma$-ray to neutron emission ratio, are summarized in table~\ref{tab:oneTableToRuleThemAll}.

%% file: g_fluence.tex
To allow LZ to estimate the $\gamma$-ray induced background of the AmLi neutron sources during calibration, we measured the gammas emerging from all three sources individually with GeIII. A simulation model was validated by comparing the simulated to the measured energy spectrum seen by the detector. This model is then used to determine the gamma-ray emission rate from the sources.

\paragraph{Gamma model development}
Almost all of the gammas emitted from the source are from the \Am decay. \Am $\alpha$-decays to various excited states of \textsuperscript{237}Np. According to the National Nuclear Data Center~\cite{nndc}, 75.449\% of these decays produce X-rays and $\gamma$-radiation. However,  $99.925\%$ of these gammas have an energy of $59$ keV or less, whose characteristic attenuation length in tungsten is 0.15 mm~\cite{nistXcom}. By design, the tungsten wall thickness of the inner and middle capsule combined is at least  4 mm, resulting in a mass attenuation factor of $3.4\times10^{11}$ or more. Hence, only the other $0.075\%$ of the gammas with energy greater than $59$ keV were simulated to save computational time. In addition to these gammas, another contribution comes from $^{7}$Li($\alpha,\alpha '$)$^{7}$Li$^{*}$ inelastic scattering that leaves the \Li nucleus in its first excited state, which decays via the emission of a 478 keV photon~\cite{PhysRev.96.389}. Further $\gamma$-radiation with 312 keV comes from the \Am progeny \textsuperscript{233}Pa decaying to an excited state of \textsuperscript{233}U. Two corresponding effective branching ratios were deduced from the AmLi source spectrum measurement and added to the existing list of branching ratios. This modified spectrum, dubbed as the AmLi gamma model, serves as the MC simulation's primary generator.

\begin{figure}[htb!]
    \centering
    \includegraphics[width=0.85\textwidth]{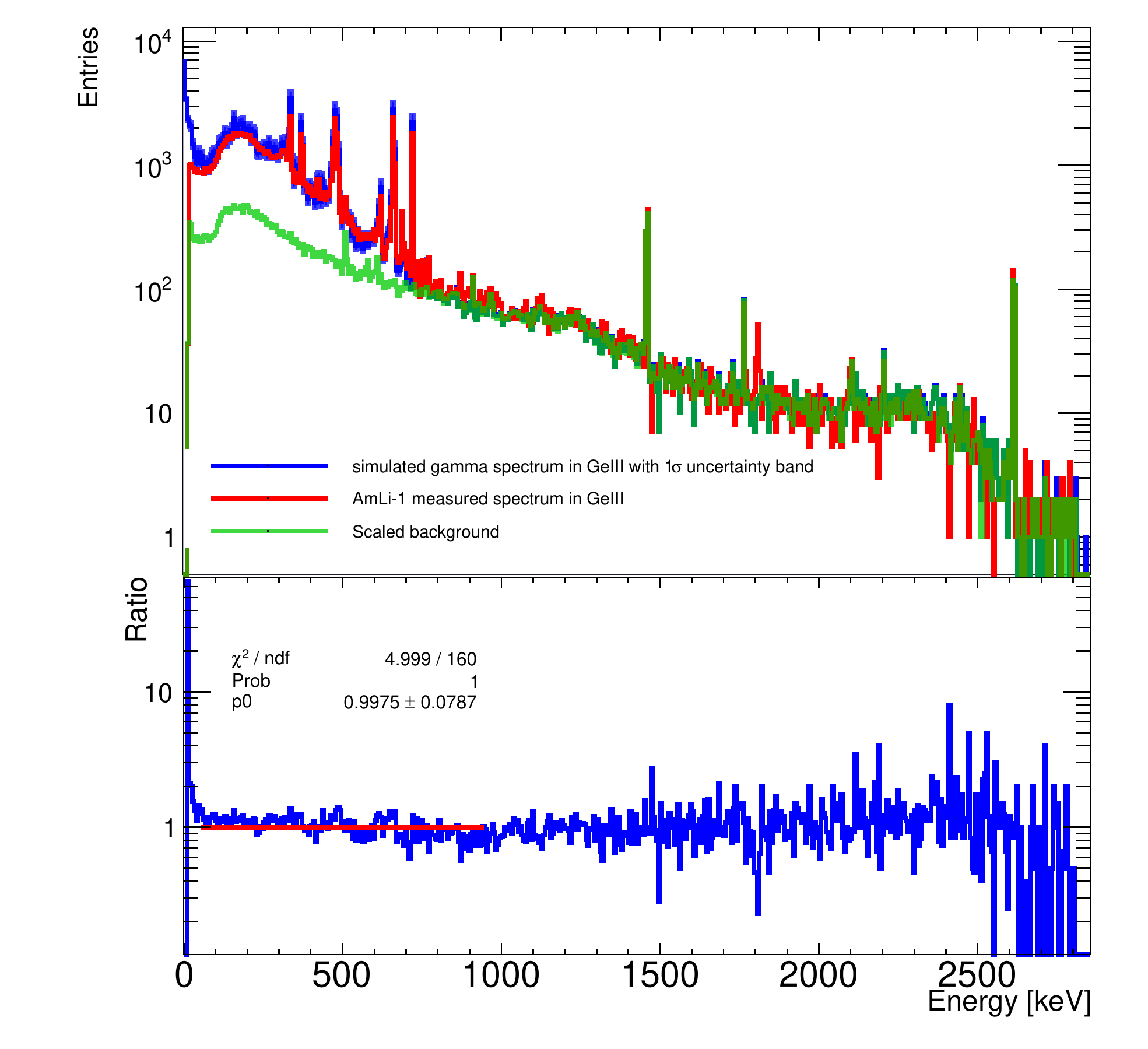}
    \caption{Simulated (blue) and measured (red) $\gamma$-spectra are superimposed for AmLi-1. A narrow 15\% uncertainty band (light blue shade) is shown around the simulated spectrum, enveloping the measured spectrum completely. For reference, the appropriately scaled detector background spectrum (green) is shown.  The bin-by-bin ratio of the simulated to measured spectrum is shown in the bottom plot for AmLi-1. The ratio is fitted with a constant between 50 - 950 keV (bottom).}
        \label{fig:simDatComparison}
\end{figure}

\begin{figure}[ht!]
     \centering  
     \begin{subfigure}[b]{0.41\textwidth}
         \centering
         \includegraphics[width=\textwidth]{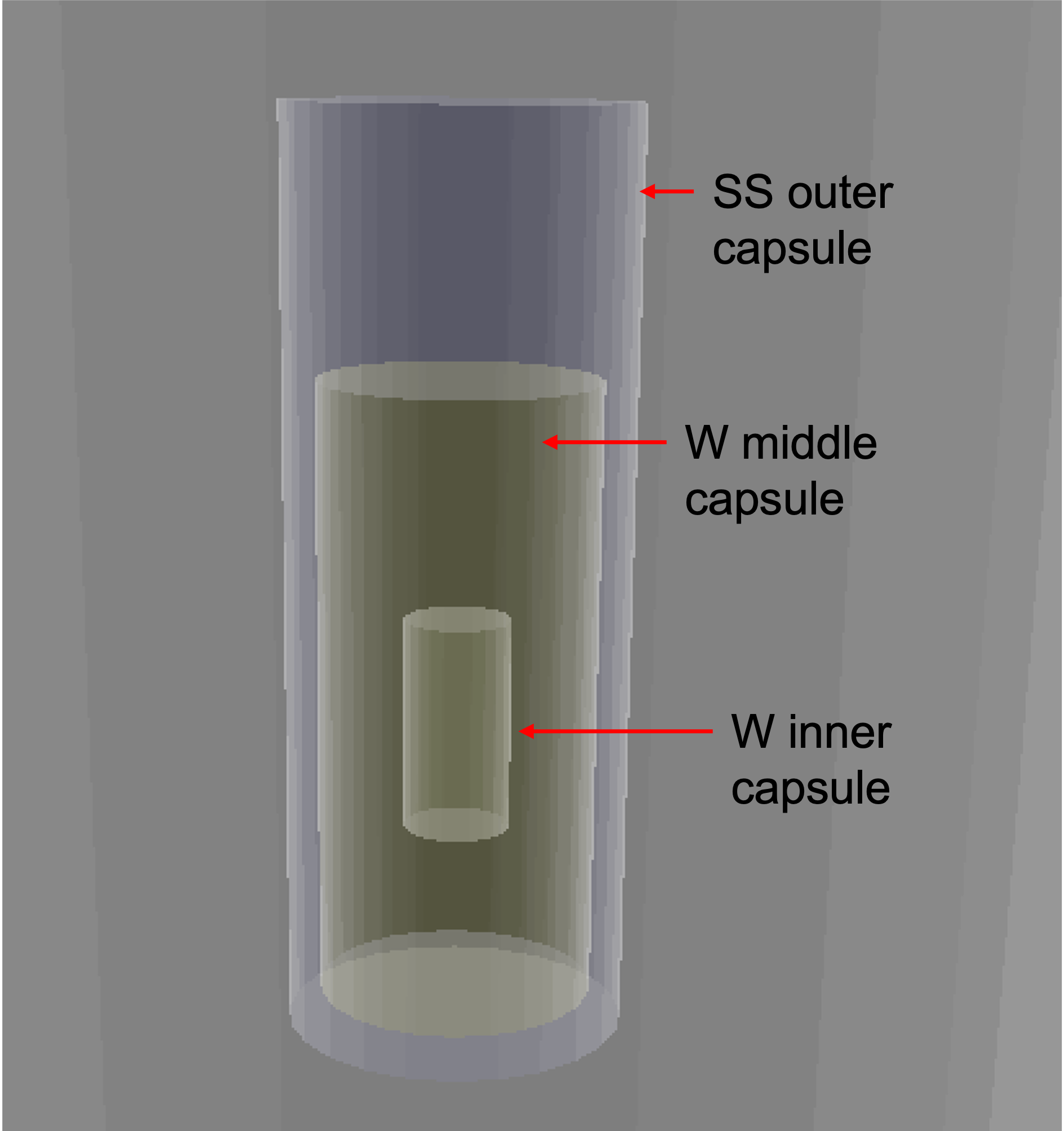}
         \label{fig:AmLisourceEncapsulations}
     \end{subfigure}
     \begin{subfigure}[b]{0.58\textwidth}
        \centering
        \includegraphics[width=\textwidth]{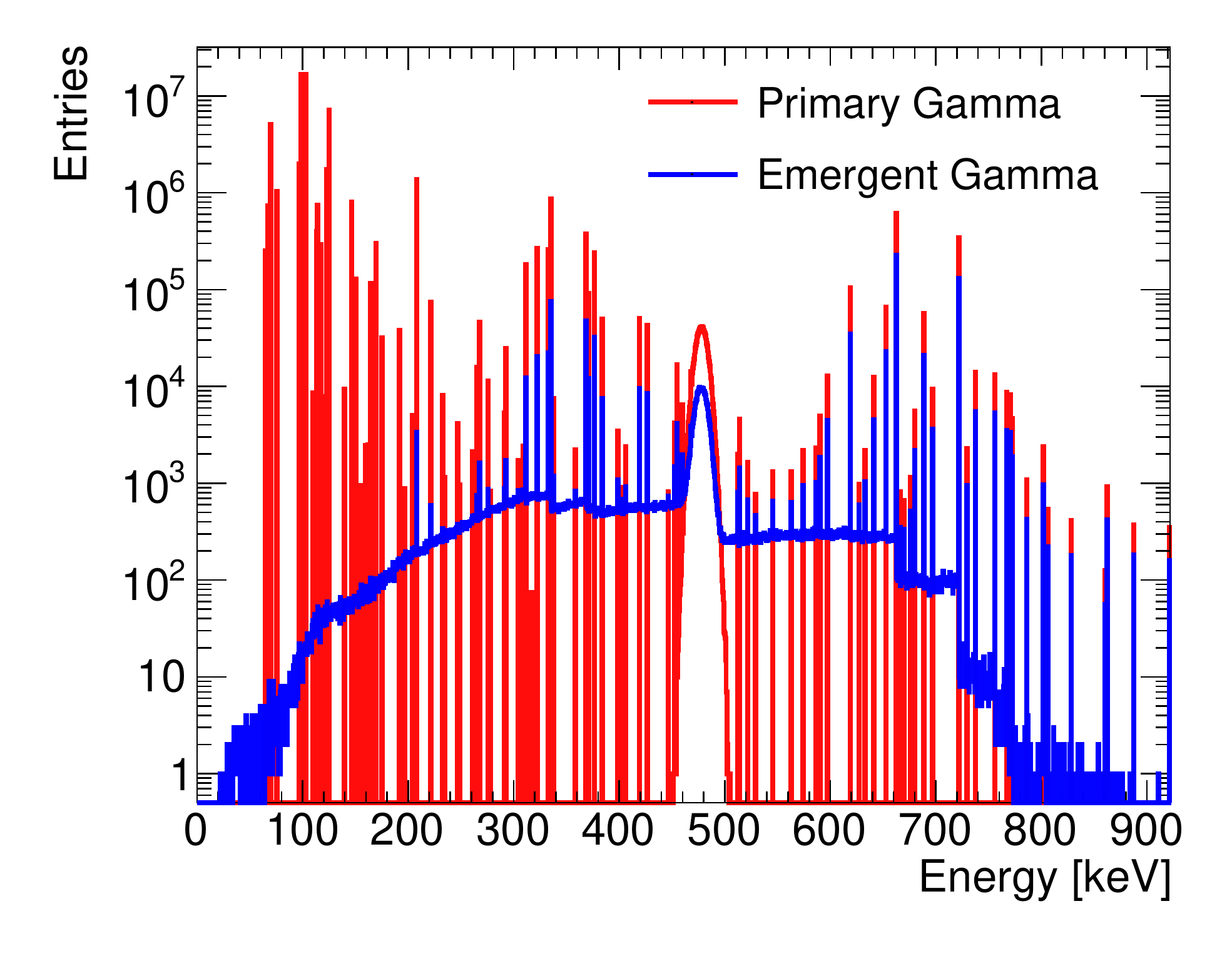}
        \label{fig:emergentGamma}
     \end{subfigure}
        \caption{Left: AmLi source encapsulation geometry used by radiation transport calculation with \textsc{Geant4}. Right: Comparison of the primary AmLi gammas (red) at the source core and the escaping gammas (blue), emerging at the outer surfaces of the source encapsulation. The red delta functions correspond to the $\gamma$-radiation implemented in the simulation. The 312 keV photons  and 478 keV gammas were implemented too. Note the Doppler broadening of the  478 keV photons (implemented by Gaussian folding using a measured spectrum), emitted by the nuclear system while in motion. The continuous part of the blue distribution corresponds to gammas Compton-scattered by the source encapsulation.}
        \label{fig:AmLiescapedGammas}
\end{figure}

\paragraph{Validation of the gamma model}
The AmLi source spectrum observed by the germanium detector is composed of primary and Compton scattered $\gamma$-radiation. The task is to unfold the detector response and scattering contribution from the data to obtain the emergent $\gamma$-spectrum as emitted by the sources. This is done by means of a \textsc{Geant4}~\cite{AGOSTINELLI2003250} Monte Carlo simulation. The primary photons are simulated, the radiation transported through the source geometry and the detector response modelled. A picture of the simulated geometry is shown earlier in the paper in figure~\ref{fig:MCforAmLiGammaMeas}, a close-up view of the source at ``remote location'' is shown in figure~\ref{fig:AmLiescapedGammas} (left). The source and detector model is validated by comparing the measured and simulated energy distributions, as shown in figure~\ref{fig:simDatComparison}.
The peaks at 312 and 478 keV have been added, as explained above. The primary $^{241}$Am gamma spectrum was generated using single gammas and statistically sampling all known branching fractions, as given by~\cite{nndc}.
As can be seen in the top part of figure~\ref{fig:simDatComparison}, the simulated (blue) and measured (red) distributions agree over the entire energy range. The appropriately scaled detector background spectrum (green) was added to the simulation. The lower part of figure~\ref{fig:simDatComparison} shows the ratio of simulated divided by observed spectrum. There is no obvious energy bias. The fit to a constant yields $0.998\pm 0.079$.

The so-verified source $\gamma$-ray emission model is shown in figure~\ref{fig:AmLiescapedGammas} (right). It was obtained by recording the frequency distribution of simulated photons at the outer surface of each AmLi source.
We call this the emergent $\gamma$-spectrum.
The source $\gamma$-ray emission rate is determined from the knowledge of the number of simulated primary gammas and counting the number of gammas crossing the outer surface of the source encapsulation. The ratio of these two numbers is multiplied by the source activity to obtain the emergent gamma emission rate of each source. The individual gamma emission rates of all three sources are tabulated in table~\ref{tab:oneTableToRuleThemAll}.

Although not prominent in the AmLi-1 $\gamma$-spectrum, a peak was observed at 871 keV for the AmLi-2 and AmLi-3 sources. 
The reaction $^{7}$N($\alpha$,p)$^{17}$O*, indicating the presence of a nitrogen contamination in AmLi-2 and AmLi-3, offers a plausible explanation of this observation~\cite{Npresence}.

\begin{figure}[ht!]
     \centering
     \includegraphics[width=0.7\textwidth]{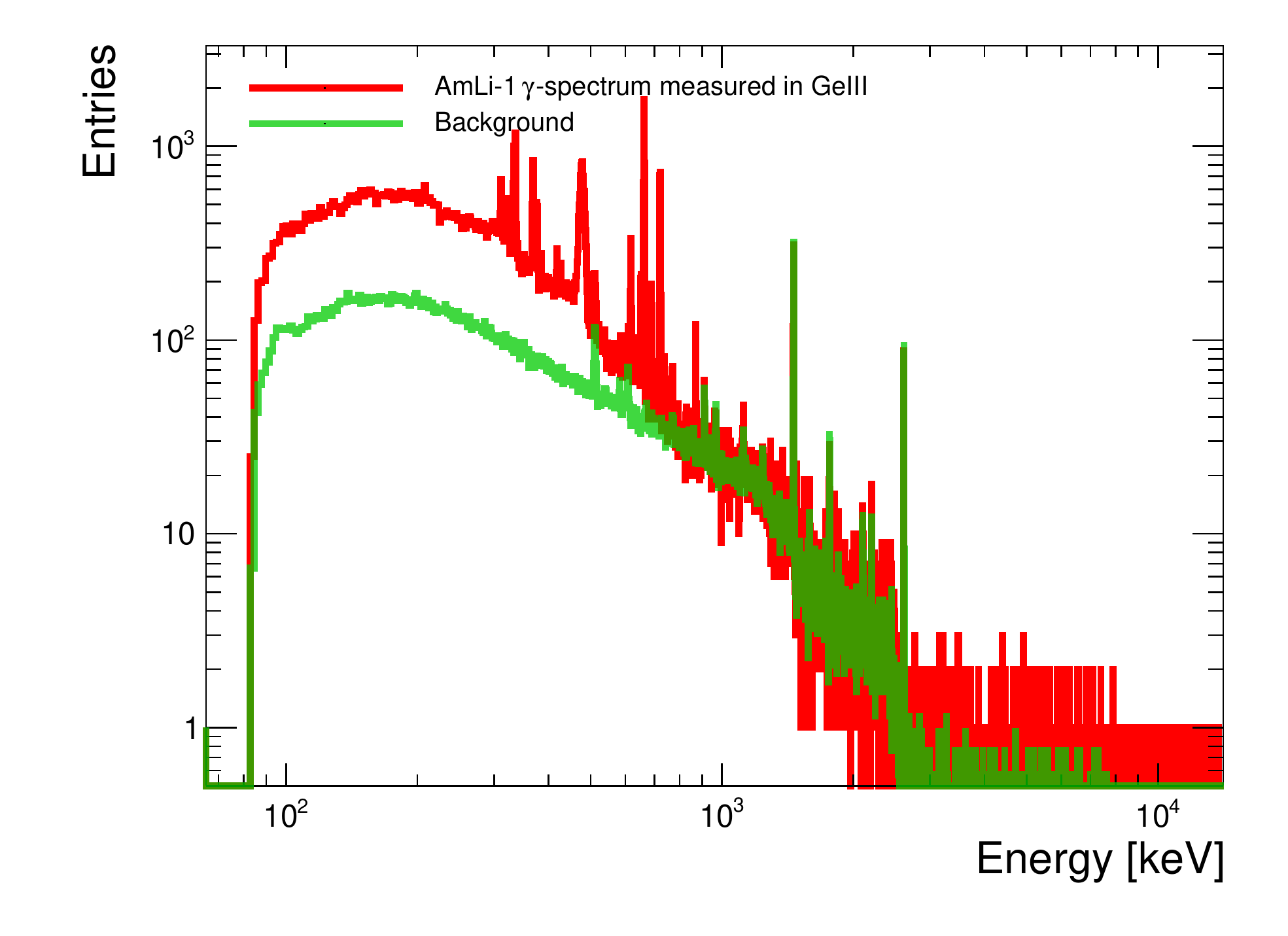}
             \caption{AmLi source high energy $\gamma$-spectrum.}
        \label{fig:AmLiHighEnergySpectrum}
\end{figure}

\paragraph{High energy gamma search}
Previous publications on AmLi sources~\cite{MOZHAYEV2021109472} identified unwanted high-energy events, created by contamination of the source with beryllium, resulting in secondary nuclear reactions. A dedicated 0.19 day long GeIII detector run with reduced amplifier gain was performed to search for the 4.439 MeV $\gamma$-radiation emitted in the de-excitation of $^{12}$C after the reaction $^9$Be($\alpha$, n)$^{12}$C$^*$. No 4.439 MeV peak was observed, as shown in the figure~\ref{fig:AmLiHighEnergySpectrum}.
A Gaussian fit to the hypothetical peak, with the centroid and width fixed to values obtained from a calibration run with an AmBe calibration source, coupled with a Monte Carlo generated detection efficiency and a gamma to neutron ratio taken from~\cite{AmBeG2N}, results in a limit of less than 3.4 AmBe neutrons/s emitted by all three AmLi sources. 

%% file: n_fluence_calibration.tex
   
\begin{figure}[ht!]
     \centering
     \begin{subfigure}[b]{0.57\textwidth}
         \centering
         \includegraphics[width=\textwidth]{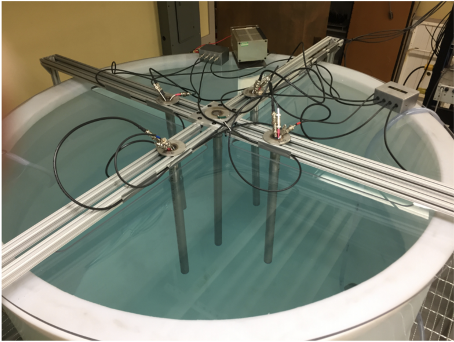}
         \end{subfigure}
         \begin{subfigure}[b]{0.40\textwidth}
         \centering
         \includegraphics[width=\textwidth]{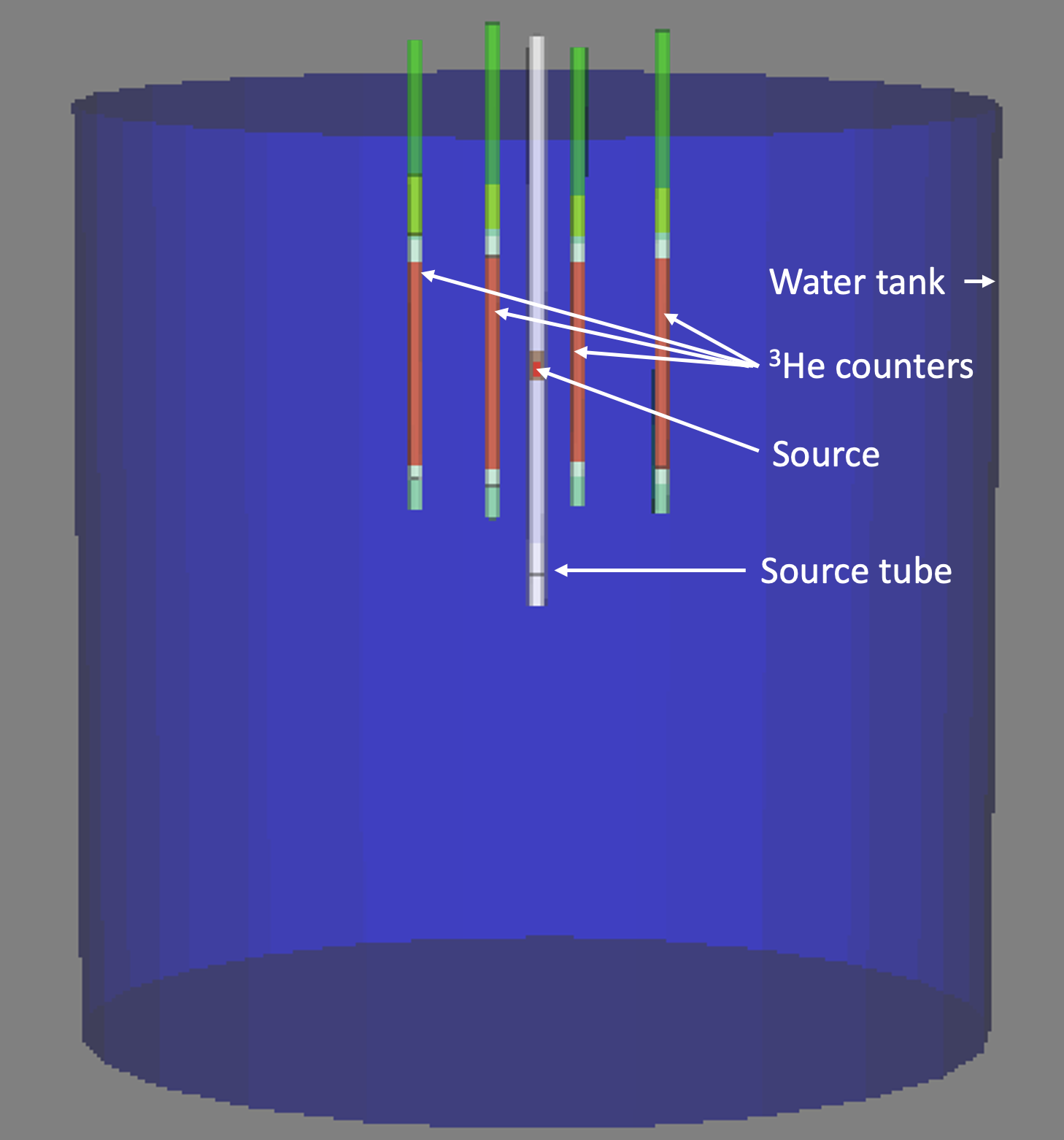}
         \end{subfigure}
         \caption{Left: $^3$He neutron detector system with four neutron counters housed in four stainless steel tubes and a central tube for the deployment of the neutron source. All tubes are submerged in water. Right: \textsc{Geant4} implementation of the detector geometry.}
         \label{fig:NDetector}  
\end{figure}

\paragraph{Experimental setup} 
The neutron emission rates of the sources were measured using four $^3$He gas-filled proportional tube counters. The model RS P4-0813-101 $^3$He tubes are from GE Reuter Stokes~\cite{reuterstokes}. 
The $^3$He tube pulses were monitored using Precision Digital Technology (PDT) modules~\cite{PDTmodules}, one PDT-20A\_HN module in combination with three PDT-10A\_HN modules. The PDT-20A\_HN contains an adjustable high-voltage generator, providing high voltage to all modules. All counters are placed in a stainless steel housing (SS304) and are suspended from a rail system into a water shielding tank. 
The rail-system allows the counters to be placed at different radial distances from the center of the water tank. At the center of the tank is a fixed stainless steel tube for neutron source deployment. For deployment, sources are attached to a stainless steel rod that can move them to six different depths inside the center tube air cavity. Water is used as a shield and for the moderation of the fast neutrons. The moderator enhances the neutron capture rate as the $^3$He thermal neutron absorption cross-section is $\sim$6000 times larger than that for fast neutrons~\cite{Rinard1997NeutronIW}. The water tank has a diameter and a depth of about 1.5 m, each. The detector setup is shown in figure~\ref{fig:NDetector}~(left).

\begin{figure}[ht!]
     \centering
     \begin{subfigure}[b]{0.49\textwidth}
         \centering
         \includegraphics[width=\textwidth]{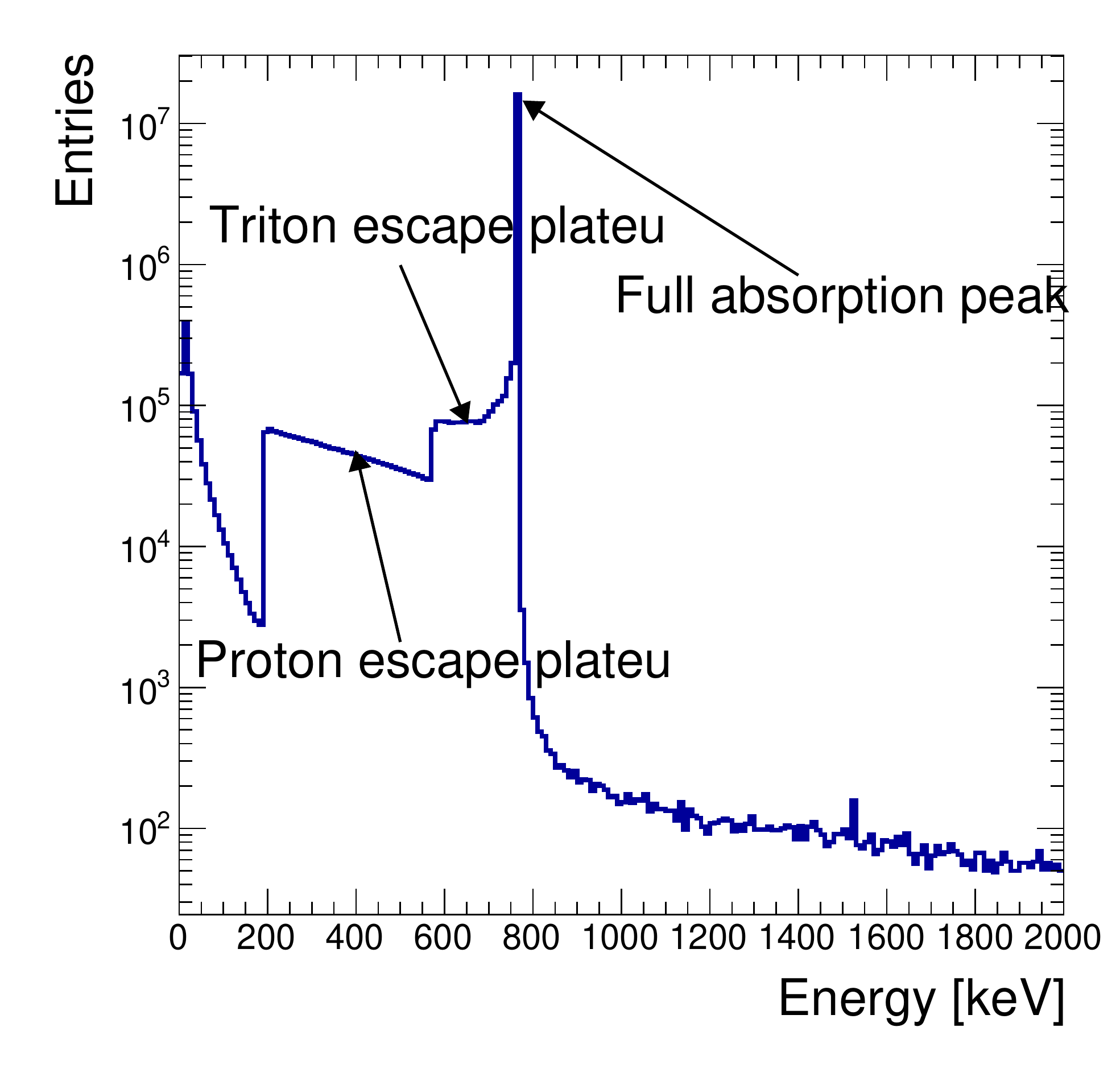}       \label{fig:cumsumToCalculateThreshold}
     \end{subfigure}
     \begin{subfigure}[b]{0.49\textwidth}
         \centering
         \includegraphics[width=\textwidth]{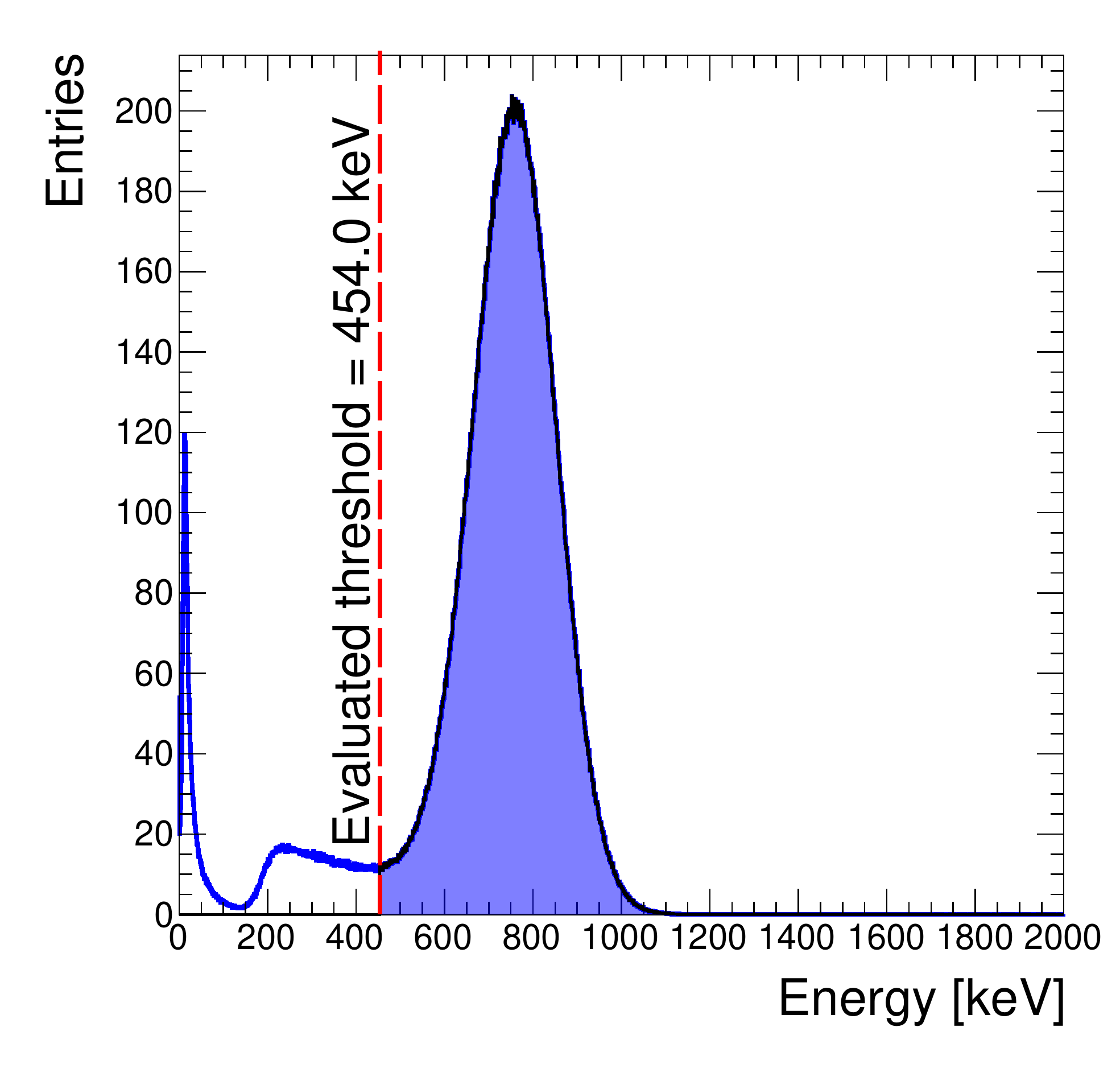}
         \label{fig:detectionThreshold}
     \end{subfigure}
        \caption{Left: Monte Carlo derived deposited energy spectrum of $^{3}$He counter 0. The detector response has not been folded with any energy resolution. Right: An example for a simulated energy spectrum of counter 0 at Position 0, folded with a 12\% energy resolution. The red dashed line represents the evaluated threshold such that the sum of the events above the threshold from MC (blue-shaded) matches the experimentally observed counts.}
        \label{fig:threshold}
\end{figure}

\paragraph{$^3$He \textsc{Geant4} simulation overview}
A response model of the setup was developed in \textsc{Geant4} (version 4.10.4.p02). The geometry of the water, source, $^3$He tubes along with their electronic modules and their stainless steel housing was implemented. Technically complex components such as electronic modules were implemented using appropriate approximations. Figure~\ref{fig:NDetector} (right) shows the setup, as contained in the \textsc{Geant4} simulation. The detectors and their holders were placed in cardinal directions with respect to the source at a distance that could be varied in the simulation. The "G4ThermalNeutron" and "QGSP\_BIC\_HP" physics lists were utilized, as recommended by \textsc{Geant4} use cases. The energy deposition characteristics of the $^3$He counters' sensitive volume were found to be consistent with that given in ref~\cite{knoll2010radiation}. Figure~\ref{fig:threshold} (left) shows the MC truth energy deposition spectrum of the $^3$He counters obtained by simulating an \AmBe source. The peak at 764 keV corresponds to the absorption of the full kinetic energy of both proton and tritium, produced after neutron capture on $^3$He. 
 The plateaus correspond to the escape of one of the reaction products from the $^3$He tube's active region, leading to incomplete energy deposition. Note the effect of the detector resolution on the spectrum, shown in the figure~\ref{fig:threshold} (right). 
 The manufacturer recommends a 10-12\% energy resolution for counters of this type~\cite{reuterstokes}. The used signal readout does not allow the digitization of pulse amplitudes. The energy response of the detectors had to be modeled using the manufacturer's recommendation. To account for this uncertainty, the change of the calculated detection efficiency under a variation of the resolution enters into the systematic error estimate.

\begin{figure}[h!]
         \centering
         \includegraphics[width=0.7\textwidth]{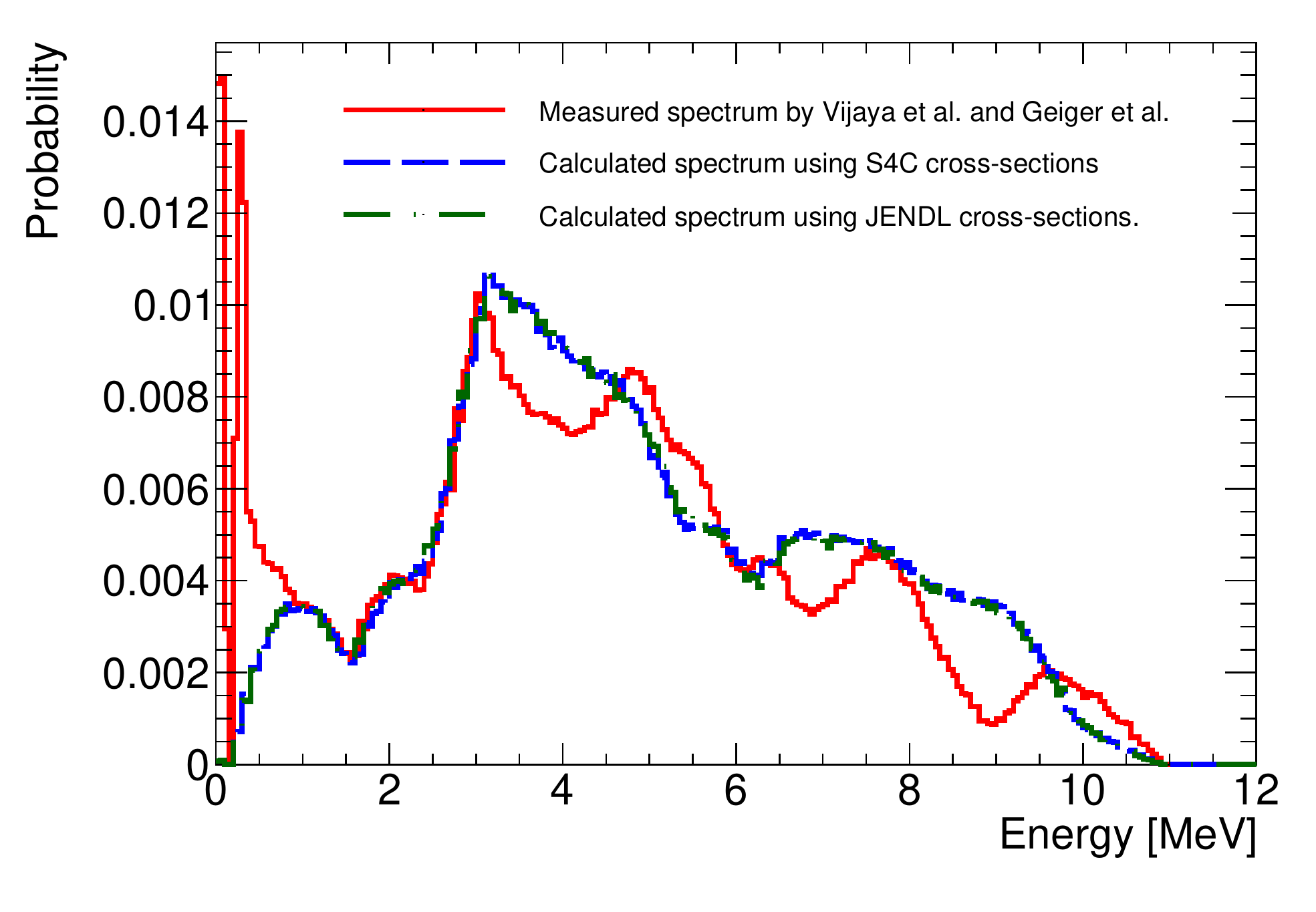}
         \label{fig:He3MonteCarlo}
        \caption{Comparison of different \AmBe input neutron spectra. The neutron spectrum derived from the S4C default cross-sections is shown in blue. The spectrum resulting from the use of JENDL-2005/AN cross-sections and a SOURCES-4C calculation is shown in green. Finally, a data-derived spectrum from~\cite{VIJAYA1973435,GEIGER1975315} is given in red.}
        \label{fig:AmBeComp}
\end{figure}

A comparison of measurement and simulation of a well-characterized \AmBe source served to verify the overall correctness of the simulation code. A comparison of measured and calculated neutron event rates, induced by a neutron emission rate-calibrated \AmBe source was used to determine the detector thresholds. 
To estimate the impact the unobserved source spectrum has on the threshold determination, calculations were performed with several plausible choices. The differences were included in the systematic uncertainty. 
One realization of the \AmBe source neutron spectrum  was generated by sampling the energy distributions given in references~\cite{VIJAYA1973435,GEIGER1975315}. 
Two more plausible choices were obtained using the SOURCES-4C (S4C) code~\cite{S4C_1}. One utilized the native cross sections of the code, the other was based on modified cross sections, taken from the JENDL-2005/AN database~\cite{JENDL}.
A comparison of the resulting AmBe neutron spectra is depicted in figure~\ref{fig:AmBeComp}.


\paragraph{Detector calibration and MC simulation tuning}
An \AmBe source with a known neutron emission rate of 488 $\pm$ 10 neutrons/s, measured by NIST, was used to calibrate the detectors and tune the Monte Carlo simulation. 
The \AmBe source-related counting rate was measured by means of a scaler, with the individual counter voltages set such as to achieve operation in the plateau region.

The detection thresholds of the counters were determined by comparing the measured rates to those derived from the simulation, normalized by the known neutron emission rate of the \AmBe source (shown in figure~\ref{fig:threshold} (right)).
During this procedure, the \AmBe source normalization was numerically varied within its uncertainty. The effect of the source rate uncertainty on the calculated thresholds was found to be less than a percent. On top of that, the detector resolution was also varied between 5 to 15\% to investigate the uncertainty it imposes on the calculated thresholds. This was also found to be less than a percent.

\begin{figure}[b]
     \centering
     \begin{subfigure}[b]{0.49\textwidth}
         \centering
         \includegraphics[width=\textwidth]{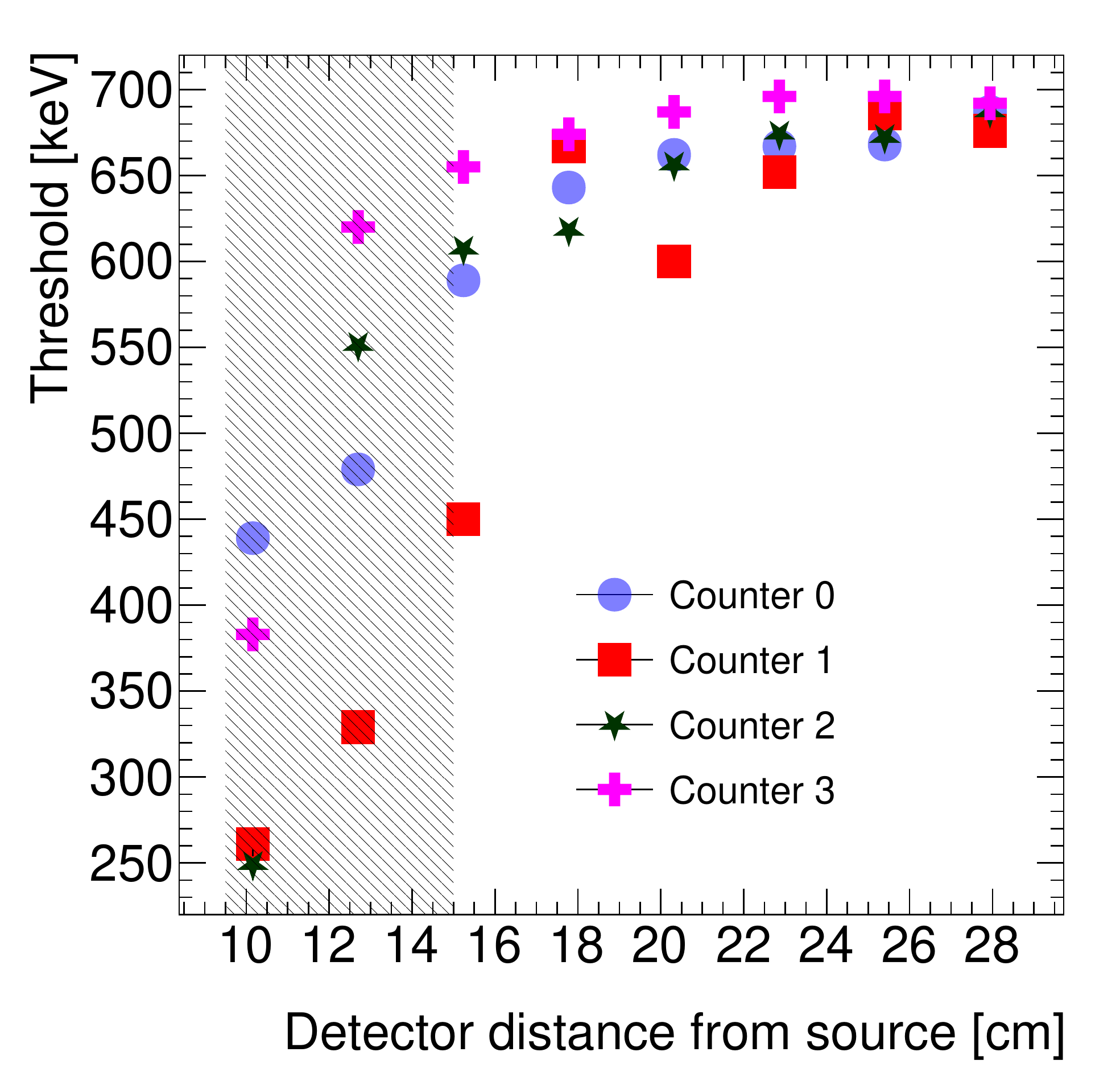}
         \label{fig:distVTh}
     \end{subfigure}
     \begin{subfigure}[b]{0.49\textwidth}
         \centering
         \includegraphics[width=\textwidth]{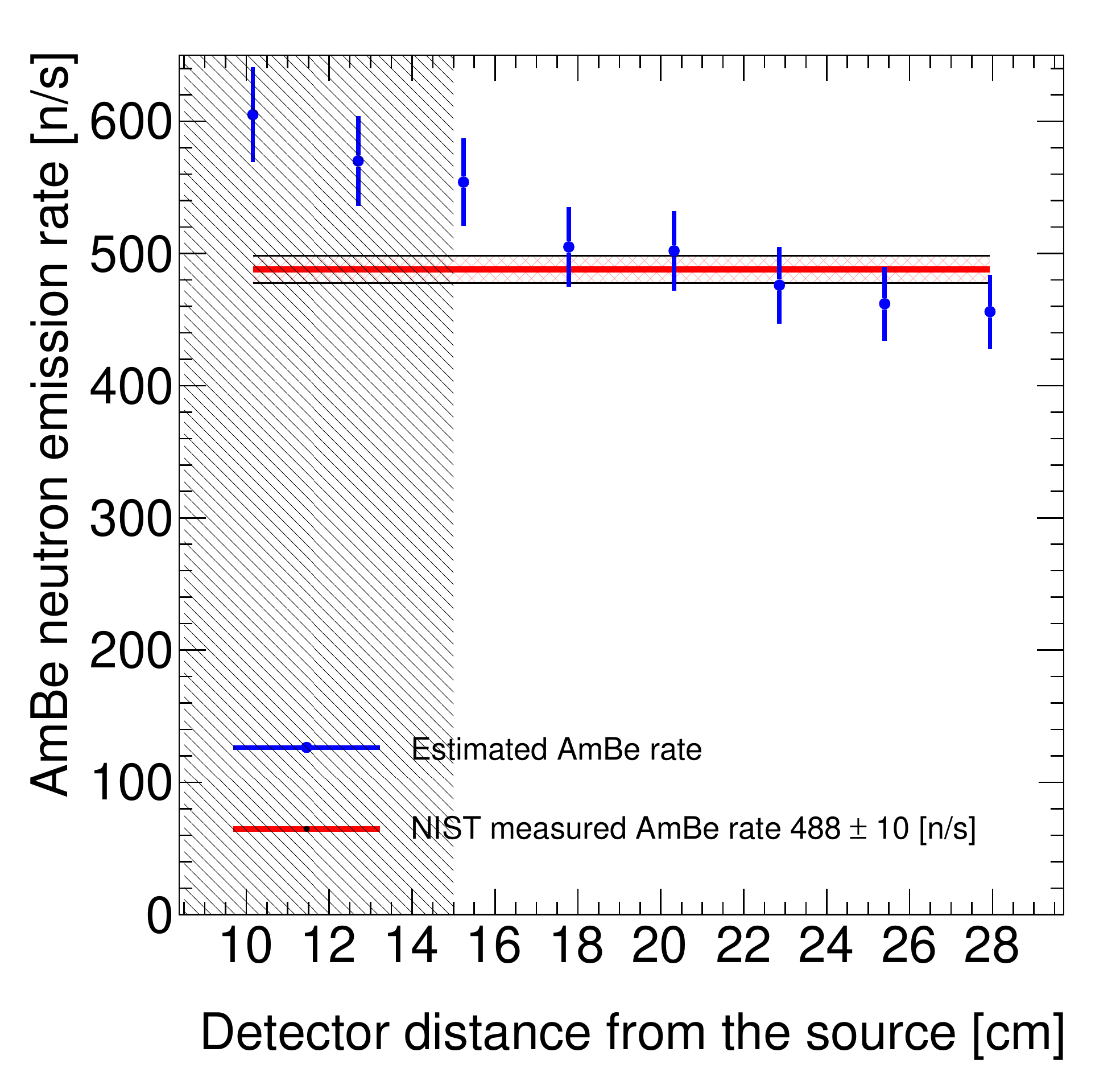}
         \label{fig:distVAmBe}
     \end{subfigure}
     \caption{Left: Detector thresholds as a function of the source-detector distance. Only measurements beyond the grey-shaded region were considered. Right: Distance dependence of the reconstructed \AmBe neutron emission rate. The NIST-measured rate (red) and its 1$\sigma$ uncertainty (red, shaded) are compared to the results derived from the tuned simulation and AmBe deployment data (blue). The result stabilizes around the true value for source-detector distances in excess of 15.2 cm.}
     \label{fig:distVsAmBeRate}
     
\end{figure}

To understand how uncertainties in geometrical source placement impact the evaluation of the thresholds, \AmBe measurements and simulations were performed for source-detector distances varying from 10 cm to 30 cm. As can be seen in figure~\ref{fig:distVsAmBeRate} (left), this variation does impact the threshold evaluation significantly.
The so-derived threshold estimates are observed to stabilize for source-detector distances in excess of 15.2 cm. 
Since thresholds are an intrinsic property of the detectors, we attribute the distance dependence seen in figure~\ref{fig:distVsAmBeRate} (left) to indicate imperfections in the treatment of thermal and epithermal neutrons, as discussed in~\cite{THULLIEZ2022166187,HARTLING201825}.
As a consequence, the thresholds were determined by averaging for distances of 15.2 cm or more. 
All measurements with the \AmLi sources were performed at source-detector distances of 15.2 cm or more.
The detector-averaged measurement-derived \AmBe source neutron emission rates obtained by applying the average thresholds are shown in figure~\ref{fig:distVsAmBeRate} (right) together with the known source activity and its error band. 

The variation in the detection thresholds of the setup, resulting from simulating different \AmBe neutron spectra was found to be less than 0.5\%. 
The various systematic uncertainties, described above, are all summarized in the thresholds uncertainties. 
The detection thresholds and their uncertainties are shown in table~\ref{tab:detector_threshold}. 
The threshold uncertainties were determined by simultaneous variation of all uncertainty components to obtain a single value. Note that the resulting error found this way is smaller than the quadratic sum of its components, indicating the presence of anti-correlation.
\begin{table}[htb!]
\caption{Detection energy thresholds of all $^3$He counters along with their uncertainty. }
\begin{center}
\input{tab/detectionThresholds}
 \label{tab:detector_threshold}
 \end{center}
\end{table}

Using the tuned simulation, a summed \AmBe source neutron detection efficiency was determined for all $^3$He detectors for various source-counter distances. 
The uncertainty of the efficiency was calculated using the method defined in~\cite{Ullrich2007TreatmentOE}. 
The 6 \AmBe neutron rate measurements, taken at distances of 15.2 cm or more (see figure~\ref{fig:distVsAmBeRate}), are combined into a single systematic uncertainty $\mathrm{\sigma_{Model}}$ by means of a $\chi^2$ statistic:
\begin{equation}
  \label{eq:chi_square}
  \chi^2 = \sum_{i=0}^{6}\frac{(R_i-R_\mathrm{{NIST}})^2}{\sigma_{R_i}^2+\sigma_{{R_\mathrm{{NIST}}}}^2+\sigma_\mathrm{{Model}}^2},
\end{equation}
where $R_i$ is the neutron emission rate at location $i$, $R_\mathrm{{NIST}}$ denotes the known neutron emission rate, $\sigma_{R_i}$ stands for the uncertainty of $R_i$, $\sigma_{R_\mathrm{{NIST}}}$ is the uncertainty of $R_\mathrm{{NIST}}$. $\sigma_\mathrm{{Model}}$ is taken to be the systematic uncertainty of the model. 
Assuming a 6\% model uncertainty, we achieve a $\chi^2$/ndf value of 6.14/6. 
This simulation model uncertainty serves as a comparison standard for a similar quantity determined for the
 \AmLi measurements described below. 

\paragraph{AmLi neutron emission rate}
Using the same procedure described above, the neutron emission rates of all three \AmLi sources were measured. 
The AmLi source measurements utilized the system of detector-thresholds derived from the \AmBe source runs. Again only measurements taken at source-detector distances of 15.2 cm or more were considered.

\begin{figure}[htb!]
     \centering
     \includegraphics[width=0.8\textwidth]{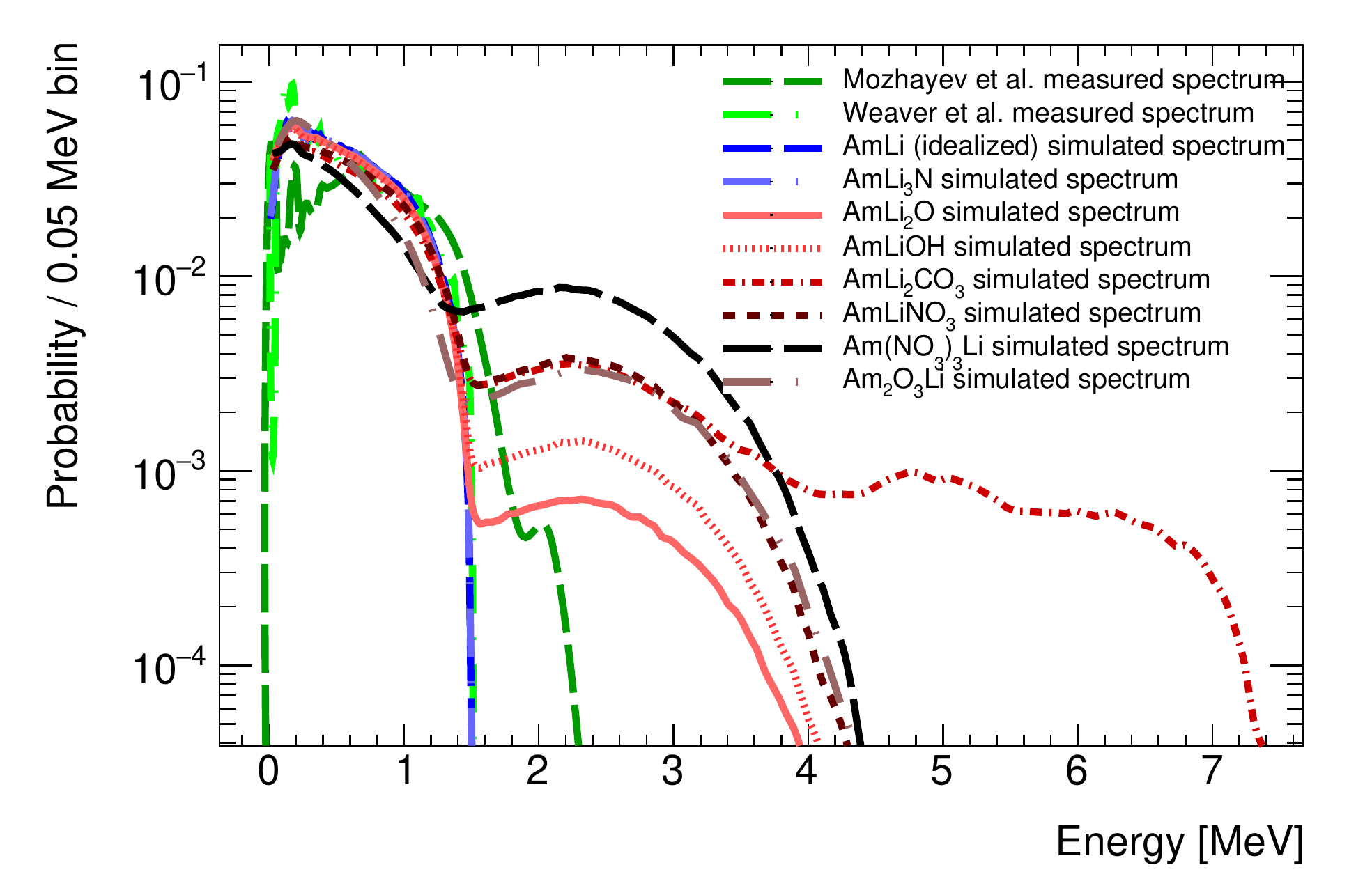}
     \caption{Comparison of simulated \AmLi neutron spectra.  Spectra derived from measurement are in green, spectra without the presence of carbon or oxygen are given in blue, and spectra with carbon and oxygen present are in red.}
     \label{fig:AmLiComp}
     
\end{figure}

As in the case of the \AmBe source, the spectrum of the neutrons emitted by the \AmLi sources is not known. 
Plausible AmLi energy spectra were taken from the literature~\cite{WEAVER1982599,MOZHAYEV2021109472} or derived from the \textsc{GEANT4} package SaG4n~\cite{Pesudo_2020,MENDOZA2020163659}. 
However, SaG4n does not allow the implementation of complex geometries; hence neutron transport through the source encapsulations for the various spectra was performed in \textsc{Geant4} with the ``ShieldingLEND'' physics list.   
In addition, the chemical form of the \Am deposit and of the \Li foils after the deposition are not known. 
To encompass all plausible source compounds (due to chemical impurities contained in the source), the following AmLi source compositions were simulated- Am-Li, Am(NO\textsubscript{3})\textsubscript{3}-Li, Am$_{2}$O$_{3}$-Li, Am-Li\textsubscript{3}N, Am-LiOH, Am-Li\textsubscript{2}O or Am-Li\textsubscript{2}CO\textsubscript{3}~\cite{LiAtmosphereInteraction,sourceContaminationPeter}. 
These composition-induced differences in the neutron energy distributions are shown in figure~\ref{fig:AmLiComp}.
The resulting differences in the neutron detection efficiency and with it the neutron emission rate are interpreted as an additional systematic uncertainty. As evident from figure~\ref{fig:AmLiComp}, the presence of americium or lithium compounds would result in a high-energy tail. Preliminary analyses from LZ indicate the potential presence of such a tail in data taken with the AmLi sources. Investigation of these events is ongoing, as is the elemental abundance of the compounds in figure~\ref{fig:AmLiComp}.

To explore the impact of source-distance variation on the neutron rate measurement, the neutron emission rate of one source, AmLi-1, was measured in detail at 5 different distances. 
The other two sources were measured at 4 distances. 
Since the neutron emission rate should not depend on the detector configuration, we interpret the standard deviation of these measured rates as a systematic uncertainty. A variability of about 5\% was found this way.   
This spatially dependent uncertainty is similar to the 6\% model uncertainty found for the \AmBe source. 
Table~\ref{tab:source_1_AmLi} shows the observed rates for AmLi-1. 
The weighted average of these rates serves as a measure for the neutron emission rate of the source. 
All systematic uncertainties and their quadratic sum are listed in table~\ref{tab:uncertaintyList}. 
The neutron emission rates of all three \AmLi sources along with our uncertainty estimate, are given in table~\ref{tab:oneTableToRuleThemAll}.
\begin{table}[htb]
\caption{The AmLi-1 neutron emission rates, determined at different source-detector distances. All uncertainties stated in this table are statistical only.}
\begin{center}

\input{tab/AmLi1NEstimation}
 \label{tab:source_1_AmLi}
 \end{center}
\end{table}

\begin{table}[htb]
\caption{Listing of systematic uncertainties of the neutron emission rates. Note that the detection threshold uncertainty is already included in the detection efficiency. Only the uncertainties below the separation line needed to be added in quadrature to the statistical error of the \AmLi neutron emission rates. 
}
\begin{center}

\input{tab/uncertaintyList}
 \label{tab:uncertaintyList}
 \end{center}
\end{table}

\paragraph{Neutron yields of the sources}
We call the number of neutrons emitted per $^{241}$Am alpha the neutron yield. 
Evaluation of this quantity allows comparison to other publications. Table~\ref{tab:oneTableToRuleThemAll} shows the yields of all three \AmLi sources. They are found to be somewhat lower than 1.5$\cdot$10$^{-6}$ n/$\alpha$, reported in~\cite{MOZHAYEV2021109472,TAGZIRIA20122395}. The chosen
sandwich geometry of our sources probably contributes here. 
Our simulation of an idealized source geometry, namely \Am alphas impinging on a flat lithium foil in the absence of carbon and oxygen, results in a yield of~(1.0~$\pm$~0.1) $\times$10\textsuperscript{-6} n/$\alpha$, twice of what is observed for our sources. 
Simulations assuming the presence of Am$_{2}$O$_{3}$, Am(NO$_{3}$)$_{3}$ also show reduced n/$\alpha$ yields.

\begin{table}[htb]
\caption{Individual AmLi source's \Am activity, gamma emission rate, neutron emission rate, gamma to neutron emission ratio, and neutron yield. }
\begin{center}

\input{tab/oneTable}
 \label{tab:oneTableToRuleThemAll}
 \end{center}
\end{table}

%% file: tab/detectionThresholds.tex
\begin{tabular}{p{3.0cm} p{1.9cm} p{1.9cm} p{1.9cm} p{1.9cm} } 
 
\hline
&Counter 0 & Counter 1 & Counter 2 & Counter 3\\ [7pt]
\hline
Thresholds [keV] &663~$\pm$~2 & 665~$\pm$~2 & 661~$\pm$~5 & 682~$\pm$~2\\

\end{tabular}

%% file: tab/AmLi1NEstimation.tex
 \begin{tabular}{p{1.5cm} p{2.0cm} p{2.8cm} p{2.0cm} p{1.8cm}} 
\hline
 \multicolumn{5}{c}{AmLi-1} \\
\hline
Detector Position [cm] & Background subtracted rate [Hz] & MC Efficiency & Neutron emission rate [Hz] & Weighted average\\ [7pt]
\hline
15.2 & 0.523~$\pm$~0.002 & 0.02871~$\pm$~0.00012  & 18.2~$\pm$~0.1 & 18.3~$\pm$~0.1\\ [7pt]
17.8 & 0.259~$\pm$~0.002 & 0.01401~$\pm$~0.00008 & 18.5~$\pm$~0.2 & \\[7pt]
20.3 & 0.116~$\pm$~0.002 & 0.00655~$\pm$~0.00003 & 17.8~$\pm$~0.3 & \\[7pt]
22.9 & 0.058~$\pm$~0.001 & 0.00304~$\pm$~0.00002 & 19.0~$\pm$~0.5 & \\[7pt]
25.4 & 0.032~$\pm$~0.001 & 0.00139~$\pm$~0.00001 & 23.0~$\pm$~1.1 & \\[7pt]
\hline

\end{tabular}

%% file: tab/uncertaintyList.tex
\begin{tabular}{p{7cm} p{2cm}}
\hline
Uncertainty Name & Uncertainty Value [$\%$]  \\ \hline
Detection threshold (contains AmBe source rate uncertainty, unknown AmBe spectrum, detector resolution uncertainty) & 0.8\\ \hline
Detector placement  & 6.1 \\
Source placement  & 0.5\\
AmLi input spectra & 2.2\\
AmLi rate spatial variability & 5.2\\
\hline
\hline
Summed systematic error [excluding detection threshold] & 8.3\\
\end{tabular}

%% file: tab/oneTable.tex
\begin{tabular}{p{1.3cm} p{2.3cm} p{2.3cm} p{2.0cm} p{2.0cm} p{2.3cm}}
\hline
Source & \Am activity [MBq] & $\gamma$-emission rate [Hz] & Neutron emission rate [Hz] & $\gamma$ to neutron ratio & Neutron yield [n/10\textsuperscript{6} $\alpha]$ \\
\hline
AmLi-1 & 31.2 $\pm$ 1.4 & 368 $\pm$ 59 & 18~$\pm$~2 & 20~$\pm$~4 & 0.59~$\pm$~0.06\\

AmLi-2 & 20.3 $\pm$ 1.0 & 239 $\pm$ 38 & 9~$\pm$~1 & 26~$\pm$~5 & 0.45~$\pm$0.04\\

AmLi-3 & 27.0 $\pm$ 1.2 & 318 $\pm$ 51 & 12~$\pm$~1 & 30~$\pm$~6 & 0.46~$\pm$~0.04\\
\hline
\hline
Total & 79.0 $\pm$ 2.0 & 925 $\pm$ 87 & 39~$\pm$~3 & & \\
\hline
\end{tabular}

%% file: Conclusion.tex
The details of the development, testing, and characterization of three low-rate \AmLi neutron sources, utilized for the calibration of the LZ dark matter experiment, are described. This includes pre-welding and quantitative leak test procedures, measurement of the \Am activities, $\gamma$-ray emission rates and spectra, the determination of neutron emission rates, and the neutron yields. The neutron emission rates of the sources have been determined with uncertainties between 8 and 11\%.